\documentclass[useAMS,usenatbib]{mn2e}
\usepackage{epsfig,subfigure,times}
\title[\textsc{Lensview}]{\textsc{Lensview}: Software for modelling resolved gravitational lens images}
\author[R. B. Wayth and R. L. Webster]{R. B. Wayth$^{1}$\thanks{E-mail:
rwayth@cfa.harvard.edu (current address); \mbox{rwebster@physics.unimelb.edu.au}}
and R. L. Webster$^{1}$\\
$^{1}$
School of Physics, University of Melbourne. Victoria 3010, Australia.}
\begin{document}

\date{}

\pagerange{\pageref{firstpage}--\pageref{lastpage}} \pubyear{2002}

\maketitle

\label{firstpage}

\begin{abstract}
We have developed a new software tool, \textsc{Lensview}, for modelling resolved gravitational lens images.
Based on the \textsc{LensMEM} algorithm, the software finds the best fitting lens mass model and source brightness distribution using a maximum entropy constraint. The method can be used with any point spread function or lens model. We review the algorithm and introduce some significant improvements. We also investigate and discuss issues associated with the statistical uncertainties of models and model parameters and the issues of source plane size and source pixel size.

We test the software on simulated optical and radio data to evaluate how well lens models can be recovered and with what accuracy.
For optical data, lens model parameters can typically be recovered with better than 1\% accuracy and the degeneracy between mass ellipticity and power law is reduced.
For radio data, we find that systematic errors associated with using processed radio maps, rather than the visibilities, are of similar magnitude to the random errors. Hence analysing radio data in image space is still useful and meaningful.

The software is applied to the optical arc HST J15433+5352 and the radio ring MG1549+3047 using a simple elliptical isothermal lens model.
For HST J15433+5352, the Einstein radius is 0\farcs525 $\pm 0.015$ which probably includes a substantial convergence contribution from a neighbouring galaxy.
For MG1549+3047, the model has Einstein radius 1\farcs105 $\pm 0.005$ and core radius 0\farcs16 $\pm 0.03$. The total mass enclosed in the critical radius is $7.06 \times 10^{10}M_{\odot}$ for our best model.

We finish by discussing issues relating to modelling of resolved lens images for this method and some alternatives.
\end{abstract}

\section{Introduction}
The discovery of the first Einstein rings at both radio \citep{1988Natur.333..537H} and optical \citep{Warren1996MN} wavelengths brought with them the promise of a new age of precision galaxy mass measurements. Strong lensing is a powerful astrophysical tool with applications ranging from measuring galaxy structural properties and evolution to cosmological parameter measurements using time delays (for $H_0$) or the statistical properties of lens samples.

Some applications of lensing, such as determining Hubble's constant ($H_0$) from time delays of multiple-image quasars, rely on accurate knowledge of both the radial mass profile of the lens mass and the location of the lens centre.
Quadruple images of lensed QSOs commonly have a range of solutions for mass slope and ellipticity \citep[e.g.][]{1994AJ....108.1156W} which cannot be resolved without additional assumptions or information.
Einstein rings can, in principle, help to solve this problem because they roughly trace the shape of the tangential critical line, hence the ellipticity of the mass distribution is well determined. In addition, the many lines-of-sight provided by Einstein rings sample the gravitational potential much more thoroughly than QSO lenses. Hence, there is a great deal of information contained in Einstein ring images.

Modelling resolved gravitationally lensed images, such as Einstein rings, is a much more complicated problem than modelling lensed QSO images. The lensed image of a resolved source is determined both by the properties of the lensing galaxy and the source brightness profile. Consequently, algorithms for modelling resolved images have gone through several stages of development.
The Ring Cycle algorithm \citep{1988A&A...191...39K,1989MNRAS.238...43K} reconstructed a source based on the assumption that the data is a true representation of the lensed surface brightness. \textsc{LensCLEAN} \citep{1992ApJ...401..461K} uses a CLEAN style algorithm to build a non-parametric source from a processed radio map. This was subsequently improved \citep{1996ApJ...464..556E} to work directly on radio visibilities rather than CLEANed radio maps. More recently, it was improved again \citep{2004MNRAS.349....1W} and applied to the lens B0218+357 \citep{2004MNRAS.349...14W} to make an accurate $H_0$ measurement.
Also recently, the `semilinear' method was developed by \citet{2003ApJ...590..673W} who noted that for a fixed lens model, determining the source model that best describes the data can be posed as a linear inversion problem and can therefore be tackled with standard matrix inversion techniques. \citet{2005PASA...22..128B,2006ApJ...637..608B} have investigated the lens modelling problem using genetic algorithms and Bayesian techniques, respectively.

An alternate approach to modelling radio images was provided by the \textsc{LensMEM} algorithm \citep[][ hereafter WNK94,WKN96]{1994ApJ...426...60W,1996ApJ...465...64W}. This approach used a pixelised source and the novel idea of a mapping matrix (relating pixels in the image and source planes) to generate a source model using an iterative approach. This algorithm is equally suitable for radio and optical data with a point-spread function (PSF) and it forms the basis of the software described in this paper.

In \S\ref{sec:how_it_works} the algorithm is reviewed and some significant improvements are described, in particular those proposed by \citet[][ hereafter SB84]{1984MNRAS.211..111S}. Also discussed is the issue of defining an acceptable goodness-of-fit with non-parametric (i.e. pixelised) sources for both radio and optical data.
In \S\ref{sec:test_cases} the software is applied to test optical and radio data, showing that for a radio map the lens model parameters can be recovered quite accurately, despite the problems associated with analysing CLEANed images.
In addition, it is shown that error estimates for the lens model parameters can be reliably made for optical and radio images using the properties of the $\chi^2$ surface.
In \S\ref{sec:examples} \textsc{Lensview} is applied to the optical arc HST J15433+5352 and the radio ring MG1549+3047.
In \S\ref{sec:conclusions} some issues relating to the software, and modelling of resolved lenses in general, are discussed and compared to some of the other methods outlined here.

Throughout this paper, we use a flat, $\Lambda$-dominated cosmology with $\Omega_m=0.3$, $\Omega_{\Lambda}=0.7$ and $H_0$=70 km s$^{-1}$ Mpc$^{-1}$.

\section{How Lensview works}
\label{sec:how_it_works}
\subsection{A review of  \textsc{LensMEM}}
In this section the basic ideas behind the \textsc{LensMEM} algorithm are reviewed and some improvements and additions found in \textsc{Lensview} are introduced.

\textsc{Lensview} is essentially an iterative method for calculating a source brightness distribution that best matches an observed lensed image for a given lens model.
\textsc{Lensview} requires a mechanism to allow the projection of a resolved source model into a lensed image while correctly preserving surface brightness.
The iterative nature of \textsc{Lensview} also requires the ability to `reverse-project' from the image plane back to the source plane (to provide feedback for the source model), an operation which is more complicated than forward projection because of the many-to-one relationship between image pixels and some source pixels.

Projection and reverse-projection are achieved by creating a `mapping matrix' which relates each pixel in the image plane to each pixel in the source plane via the lens model.
Pixels in the image and source planes represent surface brightness.
To generate the mapping matrix, image pixels (denoted $I_{ij}$) are projected back to the source plane and the area of overlap for each source pixel (denoted $f_{kl}$) is calculated.
The image pixels are divided into two triangles so that the area is calculated correctly if the pixels are folded or twisted by the projection process. This happens to pixels which overlap critical lines.
The area overlap, or `weight' (denoted $W_{ijkl}$) measures what fraction of source pixel $f_{kl}$ is covered by the projected image pixel $I_{ij}$. The weight matrix $W_{ijkl}$ has many useful properties which are repeated here from WKN96.

A source can be projected into an image using
\begin{equation}
I_{ij} = \frac{\sum_{kl} W_{ijkl} f_{kl} }{\sum_{kl} W_{ijkl}}
\label{eq:proj_source}
\end{equation}
and is typically convolved with a point spread function or ``beam'', $B_{pq}$, to generate the model image, $M_{ij}$, that a telescope would see.
The inverse magnification of an image pixel $\mu^{-1}_{ij}$ is
\begin{equation}
\mu^{-1}_{ij} = \frac{\sum_{kl} W_{ijkl} }{A_{\rm{image}}}
\label{eq:inv_mag}
\end{equation}
where $A_{\rm{image}}$ is the area of an image pixel in units of source pixel area.
The magnification in the source plane $\mu_{kl}$ is
\begin{equation}
\mu_{kl} = \sum_{ij} \left(  W_{ijkl}  / \sum_{k^{\prime}l^{\prime}} W_{ijk^{\prime}l^{\prime}} \right)
\label{eq:mag_src}
\end{equation}
where the results of the summation inside the parentheses can be performed in advance and stored.

In \textsc{LensMEM}, the software proceeds in two nested cycles: an inner loop and an outer loop. The inner loop finds the source that best fits the data (subject to the MEM constraint) for a fixed lens model. The outer loop adjusts the lens model parameters and begins the inner loop again.

Mathematically, the aim of the inner loop is to minimise the function\footnote{Note the different definition of $J$ compared to WKN96.}
$J = C - \lambda S$,  where
\begin{equation}
 C = \sum_{ij} (D_{ij} - M_{ij})^2 / \sigma^2_{ij} 
\label{eq:chisqu}
\end{equation}
 is the usual $\chi^2$ merit function
evaluated between the data, $D_{ij}$, and model image and
\begin{equation}
S = - \sum_{kl} f_{kl} \left[ \ln (f_{kl} / A) -1 \right]
\label{eq:entropy}
\end{equation}
is the entropy in the source model.
$A$ is the default pixel value for the source, or ``sky background'' (see Section \ref{sec:SB} for further discussion).

For each inner loop iteration, the source is incrementally updated $f_{kl}^{n+1} = f_{kl}^{n} + \epsilon \Delta f_{kl}$.
In  \textsc{LensMEM}, the conjugate gradient method was used so $\Delta f_{kl} = \nabla C_{kl} - \lambda \nabla S_{kl} + P_{kl}$
where
$\nabla S_{kl}$ is the gradient of the entropy\footnote{$f_{kl}$, $\Delta f_{kl}$, $\nabla C_{kl}$, $\nabla S_{kl}$ and $P_{kl}$ are \emph{images} the size of the source plane. $\epsilon$ and $\lambda$ are scalars.},
$\nabla C_{kl}$ is the gradient of the $\chi^2$ reverse-projected back to the source plane and
$P_{kl}$ is a weighted sum of previous $\Delta f_{kl}$'s as defined by the conjugate gradient technique.
Generating $\nabla C_{kl}$ is a complicated process and the details are discussed in Section \ref{sec:reverse_proj}.
Once $\Delta f_{kl}$ has been calculated, it must be scaled so that $J$ is minimised for the iteration.
To determine the scaling factor, $\epsilon$, a trial image using the source $f_{kl} +\epsilon \Delta f_{kl}$ is generated, forward projected and convolved with the PSF to calculate $\chi^2$.
Thus at each inner loop iteration, the problem is reduced to a one-dimensional line minimisation involving several forward projections to find the scale factor\footnote{This line minimisation procedure must also be iterative because the generation of the trial source, $f_{kl} +\epsilon \Delta f_{kl}$, can, in principle, generate negative source pixel values which must be corrected. If the entropy constraint is not used (i.e. $\lambda=0$), then source pixels can be negative and multiple forward projections are not necessary.}, $\epsilon$.

The inner loop stops when the minimum in $J$ is reached. At this point, $\lambda$ is adjusted and the inner loop re-started.
This process continues for up to 100 iterations.
WKN96 describe the $\lambda$-adjustment process as beginning with entropy-dominated source models (large $\lambda$ in our terminology) then decreasing $\lambda$ until a good fit is found, without providing specific details.

\subsection{Departures from \textsc{LensMEM}}

Since this is a computational method, we are interested in algorithm accuracy, stability and efficiency.
In the computational context, there are some issues with the implementation of \textsc{LensMEM} as defined in WKN96.
Firstly, there is the process of adjusting $\lambda$.
As described, the inner loop is inefficient --- effectively, a `middle' loop has been introduced where $\lambda$ is adjusted when the inner loop is making little progress, but before the lens model is changed.
At the heart of this issue is a lack of well defined stopping point, or ``good fit'' to the data. We elaborate on this in \S\ref{sec:targ_chisqu}. From an algorithmic perspective, we want to adjust the source and $\lambda$ simultaneously on the way to minimise $J$ which reduces the number of forward/reverse projections required.

Secondly, $\nabla C_{kl}$ and $\nabla S_{kl}$ are vectors that are the result of operations on completely different spaces, so adding them to form $\Delta f_{kl}$ can lead to problems such as negative source pixels. In addition, the definition of the entropy $S$ is WKN96 allows individual pixels in the source to dominate both $S$ (hence $J$) and $\nabla S_{kl}$, which affects both stability and accuracy and is undesirable.

Lastly, the fixed number of iterations of the inner loop artificially limits the ability of a lens model to reproduce the data.
Our experiments showed that models with a broad PSF (e.g. radio data) generally converge more slowly than data with a narrow PSF (e.g. HST data).
Furthermore, WKN96 state that ``with a good model, the final $\chi^2$ can typically be far smaller than $N_{\rm{pix}}$''.
This highlights the importance of knowing the target $\chi^2$ for the model.
\emph{The point of MEM is to find the most likely source given the data subject to $\chi^2$ reaching the target value.
Data with appropriate errors should not generate a $\chi^2$ significantly lower than the target even for the correct model.}
Poor lens models are never able to reproduce the data, so will converge to some $\chi^2$ greater than the target regardless of $\lambda$.
Using the Skilling \& Bryan method, good lens models converge naturally to the target $\chi^2$ because the entropy weight $\lambda$ is adjusted as part of the inner loop.
With \textsc{Lensview}, a more flexible criteria for stopping the inner loop is adopted.
If the model does not show a fractional improvement of $\sim 10^{-3}$ between iterations, or the target $\chi^2$ has been reached, the inner loop is stopped. Of course, one may want to find the best possible source given a particular lens model, in which case the target $\chi^2$ can simply be set to some impossibly low value. In that case, the entropy effectively becomes a constraint that source pixels cannot be negative.

SB84 identified these issues (and others) as general features of this kind of problem and proposed an elegant algorithm to solve the problem.
In \textsc{Lensview} we dispense with the conjugate gradient method, the $\lambda$-adjustment process and the stopping criteria of \textsc{LensMEM} and adopt SB84's general purpose algorithm to solve the minimisation problem.
In Section \ref{sec:SB} the Skilling \& Bryan method is briefly reviewed and its implementation in \textsc{Lensview} is described.

\subsection{Determining the target $\chi^2$}
\label{sec:targ_chisqu}

Determining the target $\chi^2$ is essential if meaningful comparisons between lens models are going to be made.
In the general case, the number of degrees of freedom, $\nu$, in a modelling problem is given by
\begin{equation}
\nu = N_{\rm{data}} - N_{\rm{model}}
\end{equation}
where $N_{\rm{data}}$ is the number of independent data points and $N_{\rm{model}}$ is the number of model parameters.
For large $\nu$, the target $\chi^2$ is $\nu \pm \sqrt{2 \nu}$ for a 68 percent confidence interval.
However, a pixelised source model and (potentially) broad PSF creates ambiguity in $\nu$ for the lens modelling problem.

\citet{1995ApJ...445..559K} considered the issue of degrees of freedom for modelling of CLEANed radio data and concluded that
\begin{equation}
\nu = \frac{A_{\rm{tan}}}{A_{\rm{beam}}} - m
\label{eq:kochanek_dof}
\end{equation}
where $A_{\rm{tan}}$ is the area inside the tangential caustic for the lens model, $A_{\rm{beam}}$ is the area of the beam and $m$ is the number of parameters in the lens model.
In this case, $\chi^2$ is evaluated over the entire multiply-imaged region (singly-imaged regions can be fitted perfectly with point sources and provide no constraints).
This definition is interesting because it is independent of the number of pixels (or CLEAN components) in the source plane.
It is most useful for radio rings where a large fraction of the multiply imaged region contains lensed flux.
For images where the flux is confined to a small region (usually around the tangential caustic) this is not a good definition because the many pixels that contain no flux are still counted in $\chi^2$, thus reducing the power of the statistic to differentiate between models.
Both optical (e.g. ER 0047-2808) and radio (e.g. MG0751+2716) lenses can fall into this category.

An alternate strategy for defining $\nu$ in the case of a ``thin annulus'' image is to define a mask
in the image plane and only calculate $\chi^2$ over pixels in the mask.
The mask should cover the image regions (plus a little extra so that incorrect models can be appropriately penalised) and should ensure that all images of a source pixel are included in the mask.
In this case, the mask will cover pixels both inside and outside the tangential caustic, so Equation (\ref{eq:kochanek_dof}) cannot be used.
Instead we define
\begin{equation}
\nu \equiv N_{\rm{mask}} - N_{\rm{source}} - m
\label{eq:my_dof}
\end{equation}
where $N_{\rm{mask}}$ is the number of beam areas inside the mask (or pixels for CCD data), $N_{\rm{source}}$ is the number of pixels in the source plane which map to the mask region and $m$ is the number of lens model parameters.

There are two issues that must be raised relating to the number of degrees of freedom used in the source plane.

Firstly: the entropy constraint introduces correlations in the source plane. How does this affect the number of degrees of freedom in the source? In the context of the lens modelling problem, this is not a serious issue. The entropy constraint works on the overall value of $J$, so pixels are connected only loosely via their contribution to $J$. The gradient of the entropy, $\nabla S_{kl}$, is applied to individual pixels so there is no correlation between pixels introduced by this operation. When the algorithm is used to find the best possible $\chi^2$, the source model is determined entirely by the data with the exception that source pixels cannot become $\ll A$ (or negative).

Secondly: do source pixels with small values (e.g. less than the noise) contain degrees of freedom?
It is possible to reconstruct sources having a significant number of pixels which are effectively zero, in the sense that they are well below the noise level in the image. These source pixels contribute (by definition) almost nothing to the $\chi^2$. Yet in our definition (\ref{eq:my_dof}) for the number of degrees of freedom, these pixels are counted as free parameters which artificially lowers the target $\chi^2$. Instead, a threshold, $T$, can be defined, below which all pixels are set to zero. Two models are different (with 99\% confidence) if their $\chi^2$ values differ by 6.63, so $T$ should be defined such that the overall change to $\chi^2$ is $\leq 6.63$ after setting source pixels to zero. In this way only, pixels which contribute significantly to the model image (therefore $\chi^2$) are counted as degrees of freedom.
This idea is demonstrated in \S\ref{sec:test_cases}.

For radio data, where the noise between neighbouring pixels in the image plane is correlated, problems defining a target $\chi^2$ (calculated in the image plane) remain.
The use of a pixelised source plane and the entropy constraint makes Equation (\ref{eq:kochanek_dof}) an impossible target to reach, even for the correct lens model. Our tests found this to be true even with the entropy constraint disabled.
The problems arise because the basic iterative technique (using gradients of $\chi^2$ and the mapping matrix) is fundamentally different to a CLEAN-based approach where a resolved source is constructed from many point sources.
The entropy constraint means that data values that are negative also contribute more to the $\chi^2$ than they would using CLEAN components. Since regions of negative brightness close to the real sources are a common artifact of CLEANed radio maps, this effect must be compensated for somehow.

\textsc{Lensview} does not yet work in visibility space, so the problem of defining a target $\chi^2$ for radio data remains unsolved for now although the work of \citet{2005AJ....129.1760K} in determining the uncertainties of pixels in radio maps shows promise for the future. This means that, for now, it is not possible to say whether a particular lens model is
an acceptable fit to the data using radio images. It is shown in \S\ref{sec:test_cases}, however, that lens
models can be \emph{compared} reliably using \textsc{Lensview} with radio images and that lens model parameters are
reliably recovered by the software.
For optical data, where errors in neighbouring pixels are uncorrelated, these problems do not exist.
Using optical data, lens models can be quantitatively evaluated by their ability to model the data.

\subsection{Reverse Projection}
\label{sec:reverse_proj}
WNK94's equation 8 showed how to calculate $\partial J / \partial S_l$ ($\nabla C_{kl}$ and $\nabla S_{kl}$ in our notation) in one dimension.
This equation incorporates the beam and mapping matrix for reverse-projecting $ \nabla C_{ij}$ as expected.
However, it is not completely correct for the reason discussed below.

The beam must be modified before it is applied as part of $\nabla C_{ij}$ in the image plane.
We consider for a moment only the image plane with beam-convolved model image $M_{ij} = I_{ij} * B_{pq}$ where $B_{pq}$ is the PSF or beam and $*$ denotes convolution.
Before reverse-projecting into the source plane, we must calculate $\nabla C_{ij} = \partial C / \partial I_{ij}$.
Equation 8 of SB84 showed (using one-dimensional notation with no loss of generality) that the \emph{transpose} of the matrix representing the beam must be used in calculating $\nabla C_{ij}$, which was not used in WKN94.
It can be shown, after some algebra, that the equivalent transformation for a two-dimensional beam operating
on two-dimensional images is to \emph{rotate the beam image by 180 degrees}.
Thus, WNK94 Equation 8 is correct for beams that are symmetric under a $180^{\circ}$ rotation (e.g. in radio data),
but will not be correct for asymmetric beams.

We can write 
\begin{equation}
\nabla C_{kl} = \frac{\partial{C}}{\partial{f_{kl}}} = \sum_{ij} \frac{\partial{C}}{\partial{I_{ij}} } \frac{\partial{I_{ij}}} {\partial{f_{kl}} }
\end{equation}
where
\begin{equation}
\frac{\partial{C}}{\partial{I_{ij}}} \equiv \nabla C_{ij} = -2 \frac{(D_{ij} - M_{ij})*\mathbf{B}^{T}}{\sigma_{ij}^2}
\label{eq:gradC}
\end{equation}
is the gradient of $\chi^2$ including convolution with the rotated beam, {\bf\emph{B}}$^{T}$,
and 
\begin{equation}
\frac{\partial{I_{ij}}}{\partial{f_{kl}}} = \frac{W_{ijkl}}{\sum_{k^{\prime} l^{\prime} } W_{ijk^{\prime} l^{\prime}}}
\label{eq:didf}
\end{equation}
follows from (\ref{eq:proj_source}).
So, to reverse-project $\nabla C_{ij}$ we calculate
\begin{equation}
\nabla C_{kl} = \sum_{ij} \frac{ \nabla C_{ij} W_{ijkl} } {I_{\rm{TOT},ij}}
\label{eq:rev_proj}
\end{equation}
where $I_{\rm{TOT},ij} = \sum_{k^{\prime} l^{\prime}} W_{ijk^{\prime} l^{\prime}} = A_{\rm{image}}/\mu_{ij}$ is the total area mapped to by an image pixel.

\begin{figure}
\psfig{figure=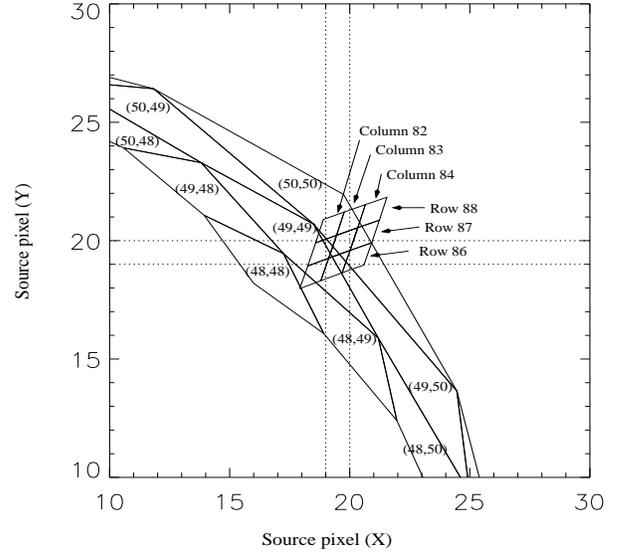,width=90mm,height=80mm}
\caption{Example tiling of the image plane onto a section of the source plane. The centre of the source plane is at x=0,y=0 (not shown).
Projected pixels from two distinct regions of the image plane are shown (shown in parentheses, or by row/column number).
Pixels around (49,49) come from very close to the centre of the lens,
whereas pixels around (83,87) are at the edge of the multiply imaged region of the lens.}
\label{fig:example_tiling}
\end{figure}

An understanding of the subtleties of reverse projection can be gained using an example.
Consider a simple elliptical isothermal-style lens.
A model is made using a $100 \times 100$ image plane and a $60 \times 60$ source plane.
The centre of the lens in the image plane is pixel $(52,52)_i$.
(Image plane
coordinates are denoted with subscript $i$ and source plane with subscript $s$ to
avoid confusion).  In Fig. \ref{fig:example_tiling} a
section of the source plane is shown with pixels from the image plane projected
onto it. There are two distinct regions of the image plane that map
to the region in the source plane around $(19,19)_s$.  The block of
projected pixels from around $(49,49)_i$ originates from very close to the lens
centre and are highly demagnified.  The block of pixels from around
$(83,87)_i$ originates from just inside the boundary of the multiply imaged
region and are slightly magnified.  Source pixel $(19,19)_s$
contributes to (and hence will take contributions from in
reverse-projecting $\nabla C_{ij}$) pixels $(50,50)_i$, $(49,49)_i$,
$(49,50)_i$, $(82,86)_i$, $(83,86)_i$, $(84,86)_i$, $(82,87)_i$ and
$(83,87)_i$.
One can see that pixels
$(83,87)_i$ and $(50,50)_i$ are the two pixels with roughly the
largest weight and $\nabla C_{kl}$ would take a roughly equal
contribution from both (from $W_{ijkl}$) with lesser contributions from the remainder.
However, pixel $(50,50)_i$ is strongly demagnified
and source pixel $(19,19)_s$ has only a small fraction of the total
area in the source plane to which it maps. The correct scheme then, is
to weight each image pixel's contribution by the fraction of the total
projected area that a source pixel contains.  This is the
same as weighting an image pixel's contribution by its magnification. This factor explicitly appears in the denominator of (\ref{eq:rev_proj}).
What about source pixels in regions of high magnification?  In such a region, there are
several (2 or more) distinct parts of the image plane mapping to the same part of the source.
In addition, several neighbouring pixels within each image region are mapped inside the same
source pixel or small region.
Again, there is a problem because there are many image pixels all contributing to the
source, none of which are down-weighted. In this case, the total
contribution to the source pixel will be too large, by a factor that
is equal to the magnification at that location in the source
plane.  To solve this problem, the overall
contribution to a source pixel must be down-weighted by $\mu_{kl}$.
This factor is accounted for by the definition of $W_{ijkl}$.

We note also that in testing the reverse projection process we tried several alternate definitions of Equation (\ref{eq:rev_proj}).
We found that some work well except in regions close to the boundary of the multiply imaged region, such as Fig. \ref{fig:example_tiling}.
In that case we found that flux was suppressed in all source pixels to which $(50,50)_i$ mapped (or the equivalent region).
The algorithm tried to compensate for this by raising the value of surrounding pixels,
causing sudden (unrealistic) jumps in the pixel values which were reminiscent of the ``glitches'' reported in WNK94.
It is possible that the glitches were caused by an incorrect reverse-projection process.
We find no such problems using reverse-projection defined by (\ref{eq:rev_proj}).
\subsection{The Skilling \& Bryan MEM algorithm}
\label{sec:SB}
SB84 provide several key insights on using iterative methods with MEM, which are summarised here.

The bulk of the computational cost in calculating $\Delta f_{kl}$
comes from convolution (via FFTs) with the beam to create the model image
and convolution with the rotated beam to create $\nabla C_{ij}$
(plus forward and reverse projection for lens modelling).
If the conjugate gradient method is used for each inner loop iteration,
a single search vector $\Delta f_{kl}$ is calculated then a one-dimensional line minimisation is performed to find the scale
factor ($\epsilon$) such that $f + \epsilon \Delta f_{kl}$ minimises $J$ for that iteration.
SB84 noted that by combining $\nabla C_{kl}$ and $\nabla S_{kl}$ into a single vector, useful information about the properties of $J$ is being discarded.
In addition, because $\lambda$ is used to combine $\nabla C_{kl}$ and $\nabla S_{kl}$ before the line minimisation is performed, it cannot be adjusted as part of the process.
SB84 found a more general and flexible model can be generated, at negligible extra computational cost,
by using a subspace of search vectors based on $\nabla C_{kl}$ and $\nabla S_{kl}$.

\textsc{Lensview} replaces the conjugate gradient based inner loop of \textsc{LensMEM} with SB84's method. The basic idea is summarised here and the interested reader is referred to SB84 for details.
For each iteration of the inner loop, a subspace is constructed spanned by three search vectors: \\
{\bf\emph{e}}$_1 = ${\bf\emph{f}}$( \nabla S_{kl})$, \\
{\bf\emph{e}}$_2 = ${\bf\emph{f}}$(\nabla C_{kl})$ and \\
{\bf\emph{e}}$_3 = ${\bf\emph{f}}$(\nabla \nabla C_{kl} )${\bf\emph{f}}$( \nabla S_{kl}) / | \nabla S_{kl}| - ${\bf\emph{f}}$(\nabla \nabla C_{kl}) ${\bf\emph{f}}$(\nabla C_{kl}) / |\nabla C_{kl}|$. \\
Note that both of the original search directions, ($ \nabla S_{kl}, \nabla C_{kl} $), are weighted
by {\bf\emph{f}}, the ``entropy metric'', (pixel by pixel multiplication, not a dot product) which is necessary and useful, but will not be discussed here.
$\nabla \nabla C_{kl}$ is simply the reverse-projected per-pixel variance image after convolution with the beam \emph{and} the transformed beam and is constant throughout the inner loop. The subspace is simply a model of the local properties of $J$ and depends only on {\bf\emph{e}}$_{\mu}$ and $\lambda$. 

The goal of the process now is to find {\bf\emph{x}}$^{\mu}$ such that {\bf\emph{f}}$ + $
{\bf\emph{x}}$^{\mu}${\bf\emph{e}}$_{\mu}$ minimises $J$ in the subspace, subject to
some sensible constraints on how much the source can be changed in a
single iteration. In addition, because $\nabla S_{kl}$ and $\nabla C_{kl}$
are not combined into a single search direction, $\lambda$ can be
adjusted as part of the same process. Indeed, for each iteration of
the Skilling \& Bryan algorithm, the minimum in $J$ (within the subspace)
always represents the maximum entropy solution for that iteration.
In principle, {\bf\emph{x}}$_{\mu}$ and $\lambda$ can be solved for algebraically, however constraints on how much the source can be sensibly changed in an iteration mean a more complicated process is required.
The actual process for determining $\lambda$ is via the ``$\alpha$-chop'' algorithm described in SB84, which does not use any projections or convolutions. Once $\lambda$ is determined (per iteration), {\bf\emph{x}}$^{\mu}$ is calculated, a new source generated and the next iteration begins.

The definition of entropy has also been adopted from SB84 (Equation \ref{eq:entropy}), which incorporates
a default source value, $A$. This is the value that a source pixel will take if there are no data constraining it.
Introducing $A$ makes $S$ a pseudo-dimensionless quantity in the sense that any source pixel which has a value
equal to $A$ will contribute zero to the entropy, regardless of the units of measurement of $f_{kl}$.
In contrast, the definition used by WKN96, $S = \sum_{kl} f_{kl} \ln f_{kl}$, does depend on the units of measurement whereas the $\chi^2$ statistic does not.
In practice this unit difference can be absorbed into the value of $\lambda$,
however pixels with values $\ll 1$ are strongly prohibited by the WKN96 definition of entropy where
$\partial S / \partial f_{kl} = -1 - \ln f_{kl}$.
The SB84 definition of entropy has $\partial S / \partial f_{kl} = \ln A - \ln f_{kl}$ which tolerates values that are not $\ll A$.

The need to define $A$, the default source value, highlights the importance of understanding
the noise characteristics of the data. This point has been stressed many times in the context of the MEM
and is just as important for \textsc{Lensview}. The noise properties of the data affect the results in the calculation
of $C$, the default source value and in the calculation of $\nabla C_{ij}$ and $\nabla \nabla C_{kl}$.
A suitable value for $A$ is roughly 1/2 the value of the background noise in the image. The results of modelling are not very sensitive to $A$ as long as it is the same magnitude as the noise.

\textsc{Lensview} is implemented such that it can use the \textsc{LensMEM}-style conjugate gradient approach or the SB84 algorithm.
Extensive testing has shown that the SB84 algorithm finds better sources\footnote{This is due mostly to the difference in definition of entropy and to a lesser extent the definition of search directions ($\nabla f$ vs $e_{\mu}$). If \textsc{Lensview} used WKN96's definition of entropy, the results should be the same in principle.} (both smoother and better $\chi^2$) and is much faster than the conjugate gradient approach. This is because the multiple forward projections to find $\epsilon$ in each iteration of \textsc{LensMEM}'s inner loop are removed. The conjugate gradient approach is useful if one wants to disable the entropy constraint by setting $\lambda=0$.

\section{Test cases}
\label{sec:test_cases}
There are two main goals when testing the software. Firstly: can the software recover the lens model parameters of a known lens, and with what accuracy? Secondly, is the source reconstruction accurate?
In this section, the software is applied to simulated optical and radio data with these two goals in mind. While working towards the high-level goal of answering the two questions above, we must also consider the issues of source plane pixel size, the overall size of the source plane and the number of degrees of freedom used in the source model.

In both tests, the data is processed in image space. For optical data, the analysis is straightforward using the data and a pixel variance image.
For the radio data, the image is not directly measured by the telescope, hence the analysis must begin with understanding what effects, if any, are introduced by using processed images.

The focus of each section will be different because the typical lensing scenario is quite different between (say) a lensed high redshift galaxy (typically $\sim 0.1$ arcsec in size) and a lensed radio lobe ($\sim 1$ arcsec in size).

For a lensed galaxy, the object is of comparable size to the tangential caustic and usually highly magnified into a thin ring or series of arcs (e.g. the lenses ER0047-2808 and the near-IR images of QSO host galaxies). The high magnification reveals structure in the source that would be unresolved if not for the lensing and is manifested through brightness variations in the ring/arcs. This structure must be accounted for in the source model so the spatial resolution of the source plane must be higher than the measured image. The overall size of the source plane, however, is small.

For a lensed radio lobe, the source is much larger than the tangential caustic, hence the overall magnification is modest. The object is lensed into a ring with surface brightness that varies, but is relatively constant around the tangential critical line. Hence it is not likely that source pixels need to have higher spatial resolution than image pixels (which are themselves typically 1/3 of the beam width). In addition, much of the image will be found away from the high magnification region but still multiply imaged. The source plane in this case needs to be large.

\subsection{A test with simulated optical data}
\label{sec:test_optical}

In this section, \textsc{Lensview} is tested with simulated optical images. The images are designed to test the discriminating power of the data for images with significantly different sizes.
The issues that specifically need to be addressed are: the optimum size of pixels in the source plane, the accuracy of lens model parameters, the distinguishing power of the image between different (non-isothermal) models and the overall goodness-of-fit of any particular model.

\begin{figure*}
\centering
\hspace{-10mm}
\includegraphics[scale=0.7]{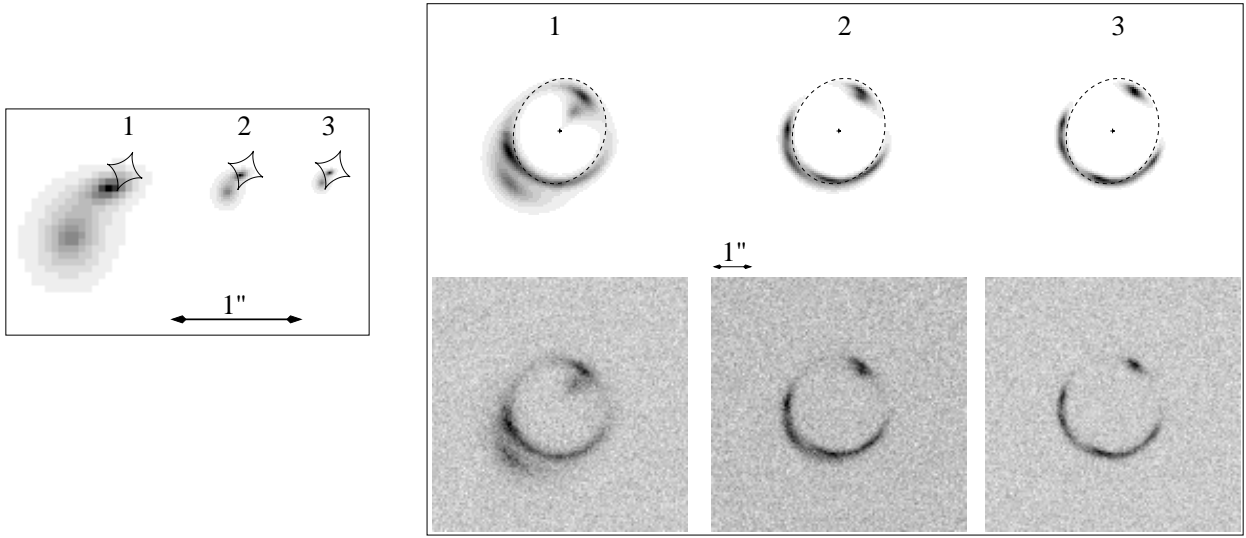}
\caption{The source and model images (noiseless and noisy) used in the optical data simulation. The source is shown with the caustic to highlight its position relative to the lens centre. The noise-free image is shown with the tangential critical line. All images have a linear greyscale stretch.}
\label{fig:fake_opt_rings}
\end{figure*}

Three model images were created using a two-component source. The source and images are shown in  Fig. \ref{fig:fake_opt_rings}.
The source is designed to have two regions of substantially different size and different peak brightness.
The source was offset from the lens centre and projected into an image with 0.05 arcsec pixels using a softened power-law elliptical mass distribution (SPEMD) \citep{1993ApJ...417..450K,1998ApJ...502..531B} lens model with non-isothermal mass slope. Each of the model images has the same lensing properties, it is only the relative size of the source that has changed. The SPEMD is defined by surface mass density
\begin{equation}
\kappa = \kappa_0  (x^2 q + y^2/q + r_c^2)^{-\beta/2}
\label{eq:SPEMD}
\end{equation}
where $\kappa_0$ is the central density, $q$ is the axis ratio, $r_c$ is the central core radius, $\beta$ is the mass slope parameter and the major axis of the system is rotated by $\theta$ degrees. The mass slope allows the SPEMD to model a range of mass profiles from constant density ($\beta=0$) to isothermal ($\beta=1$) to a modified Hubble profile ($\beta=2$). In practise, we scaled $\kappa_0$ such that  $b= \kappa_0 \frac{8 \pi G}{c^2}\left( \frac{2-\beta}{q} \right) ^{\beta/2}$ so that a single parameter, $b$, is used for the mass scale factor which represents the Einstein radius of the lens.
The parameters used to create the simulated image were $b$ = 1.2 arcsec, $q=0.82$, $r_c=0.05$ arcsec, $\beta=1.16$ and $\theta=68^\circ$.

A PSF was generated for the HST ACS instrument using the F555W filter with the \textsc{TinyTim} \citep{1993adass...2..536K} software. This PSF was convolved with the simulated image. Next, random noise was added to the image including contributions from CCD read noise, sky and signal shot noise. (The effects of lensing galaxy subtraction have been ignored here and it is assumed that the lensing galaxy has been subtracted perfectly. In addition, it is assumed that the lens galaxy light does not add significant noise to pixels in the region of the image.)
The overall signal-to-noise ratio was chosen to be similar to ER0047-2808 with the peak signal in the simulated image approximately 10 times the background noise level. The sources, PSF-convolved images and noisy images are shown in Fig. \ref{fig:fake_opt_rings}. The model images are labelled 1-3 as shown.

\subsubsection{The optimal source pixel size}
\begin{table}
\centering
\begin{tabular}{c|ll}
Ratio & $\chi^2$ & Comment \\ \hline
1 & 1266.6 & Source components not resolved.\\
 &         & Poor image reconstruction. \\
2 & 1146.5 & Barely resolved source. \\
 &         & Improved image reconstruction. \\
3 & 1086.9 & Resolved bright regions in source. \\
4 & 1055.8 & Source becoming noisy \\
5 & 1024.6 & Source very noisy
\end{tabular}
\caption{Best $\chi^2$ that could be reached using different image-to-source pixel size ratios for simulations of optical image 3. The $\chi^2$ was evaluated over a mask of 1234 pixels surrounding the image.}
\label{tab:opt_srcsize_res}
\end{table}

When reconstructing a source, pixels in the source plane need only be small enough to explain the smallest features in the image. In this section we investigate the best pixel size for image 3 because it has the finest detail.

An estimate of the smallest required pixel resolution can be made by examining the image data. The arc crossing the tangential critical line has two bright regions on each end. These regions are resolved into small arcs about 3 pixels long. Given that these images are in a high ($\sim 10$) magnification region and all the magnification in isothermal-style lenses comes from tangential stretching of the image, the size of the structure in the source must be $\sim 3/10$ the size of an image pixel. Thus, a first guess for a suitable source pixel size is 0.3 times the image pixel size, i.e. an image-to-source pixel resolution ratio of 3.33.

To test both the ability of the software to reproduce the data and the quality of the source reconstruction, image 3 was analysed with source plane pixels 1, 1/2, 1/3, 1/4 and 1/5 the size of the image pixels (i.e image-to-source pixel ratio of 1, 2, 3, 4 and 5 respectively). A mask of 1234 pixels surrounding the arcs was defined and the image was processed using the known (correct) lens model parameters. The source plane overall size was chosen to fit the reconstructed source with a little extra room.

\begin{figure*}
\centering
\includegraphics[scale=0.55]{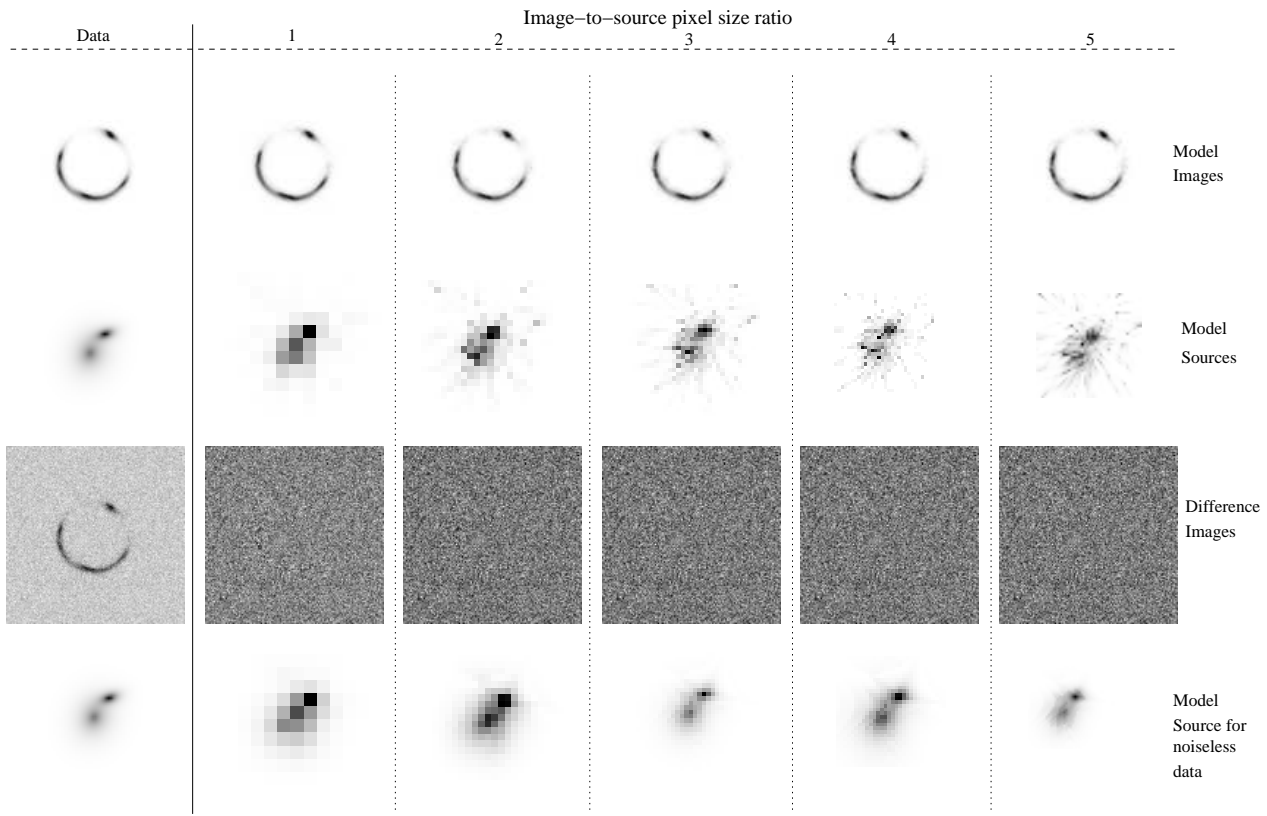}
\caption{Results for reconstructions of simulated optical data (test image 3) with different source pixel sizes. The source planes have been scaled to the same angular size. All images have a linear greyscale stretch. The image on the left in the row ``difference images'' is the actual image, including noise, used in the modelling.}
\label{fig:opt_sim_source_sizes}
\end{figure*}

Results of the modelling are shown in Table \ref{tab:opt_srcsize_res} and graphically in Fig. \ref{fig:opt_sim_source_sizes}. The quality of the model image improves dramatically for a pixel ratio of 2 (compared to 1) and again for a pixel ratio of 3. For pixel ratios of 3 and above there are no discernible residuals between the data and model image.
The meaning of the $\chi^2$ values depends on the number of degrees of freedom in the source --- a topic that is discussed later. However, it is clear that a pixel ratio of 1 produces a poor model image.
For higher resolution source planes, the source reconstruction becomes very noisy because distinct regions of the source are able to model individual noisy pixels in the image. Hence they are affected by the noise in the image. At medium resolution (ratio of 2-3), the source is detailed enough to be able to reconstruct the image but the pixel mapping overlap from neighbouring pixels helps to smooth out the source.

From this qualitative analysis, it is clear that an image-to-source pixel scale ratio of at least 2 is required to reproduce the data and a ratio of 3 is optimal for this data. This agrees well with the initial guess for a ratio of 3.33. Also shown in Fig. \ref{fig:opt_sim_source_sizes} are the source reconstructions for the noiseless (but still PSF-convolved) image. In that case, the reconstructions match the original source very well and are limited only by the finite resolution of the image.

\subsubsection{Model goodness-of-fit and source degrees of freedom}
\begin{table*}
\centering
\begin{tabular}{c|crcccccc}
Source&          &Target&     &     &                       & Target & New     & \\
 size & $\chi^2$ &(old) & OK? & $T$ &$N_{\mathrm{s}} > T$ & (new)  &$\chi^2$ & OK? \\ \hline
10x10 & 1353.6   & 1130 & N   & 1.0 & 99  & 1131 & 1353.5 & N \\
15x15 & 1154.2   & 1005 & N   & 1.0 & 176 & 1054 & 1154.3 & Y \\
20x20 & 1119.9   & 830  & N   & 1.0 & 215 & 1015 & 1123.3 & Y \\
25x25 & 1100.5   & 605  & N   & 1.0 & 232 & 998 & 1106.4 & Y \\
30x30 & 1088.0   & 330  & N   & 1.0 & 242 & 988 & 1094.5 & Y
\end{tabular}
\caption{Results for differently sized source planes that have source pixels $< T$ set to zero. Column 1 gives the size of the source plane in pixels. Column 2 is the best $\chi^2$ attainable with that source plane. Column 3 is the target $\chi^2$ using equation (\ref{eq:my_dof}). Column 4 lists whether the $\chi^2$ is within a 99\% acceptability range. Column 5 lists the threshold value used for the source plane. Column 6 lists the number of pixels in the source that are above this threshold. Column 7 lists the revised target $\chi^2$ using equation (\ref{eq:my_dof_new}). Column 8 lists the $\chi^2$ for the source with pixels below $T$ set to zero. Column 9 lists the acceptability of the new model with the revised target.}
\label{tab:opt_src_thresh}
\end{table*}

Using Equation (\ref{eq:my_dof}), an acceptable range for $\chi^2$ can be defined for the simulated data.
The equation relates the degrees of freedom used in the source plane directly to the target $\chi^2$.
In this section, this idea is explored further.

\begin{figure}
\centering
     \includegraphics[scale=0.9]{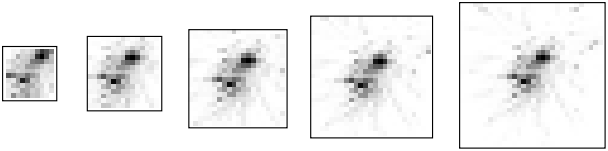}
\caption{Reconstructed sources for the simulated optical data with differently sized source planes. The sources are 10x10, 15x15, 20x20, 25x25 and 30x30 from left to right. For all images, the image-to-source pixel size ratio is 3.}
\label{fig:opt_diff_source_sizes}
\end{figure}

Using an image-to-source pixel size ratio of 3 and a variety of source plane sizes, the best possible $\chi^2$ the model could achieve (subject to the entropy constraint, of course) was found, again for test image 3. In each case the source plane is positioned so the reconstructed source is as close to the centre as possible to minimise any potential edge effects. The resulting reconstructed sources are shown in Fig. \ref{fig:opt_diff_source_sizes}.

It is clear from these images that once the source plane is large enough (approximately $15 \times 15$ pixels in this case), most of the remaining pixels in the image do not contribute anything to the model.
Using the pixel thresholding scheme presented in \S\ref{sec:targ_chisqu}, the simulated data was analysed again with pixel ratio 3 and the 1234 pixel mask using a 99\% confidence threshold for $T$. Hence, $\chi^2$ was permitted to change by up to 6.6.
Table \ref{tab:opt_src_thresh} shows the best $\chi^2$ that can be reached using different sized source planes and the interpretation of the model goodness-of-fit both with and without the thresholding scheme.

The table shows that large source planes have a minimal effect on improving the model but they do help to clearly identify the location and extent of the source. (For very large source planes, the pixels will map outside the pixel mask so by definition do not contribute to $\chi^2$.) Hence, the simple definition (\ref{eq:my_dof}) for source degrees of freedom is oversimplified for all source planes which have many low-value pixels.
The threshold-based definition means the results do not depend on the particular size of the source plane (unless it is too small), which is the result one intuitively expects.

Using the threshold based definition for $\nu$,
\begin{equation}
\nu \equiv N_{\rm{mask}} - (N_{\rm{source} > T}) - m
\label{eq:my_dof_new}
\end{equation}
it is clear from Table \ref{tab:opt_src_thresh} that all reasonably sized source planes are able to produce a model which is statistically acceptable. Only the $10\times10$ model was not within 99\% confidence because it was too small to contain the source. For large source planes, regions of the plane that map outside the image mask are not affected by the data hence will become the default pixel value, $A$. These pixels then do not affect the fit at all because they have zero contribution to $J$ and $\nabla S_{kl}$.

Given that the noise properties of the image and lens model are known exactly, it is very reassuring that the thresholding scheme produces the expected results.

\subsubsection{Errors on fitted parameters}
The focus of testing thus far has been on the properties of the source and source plane using a fixed (correct) lens model. Obviously the lens model parameters are not known with real data and the major outcome of modelling is a set of measured lens model parameters.
Thus, the next task is to quantify the uncertainty in the lens model parameters themselves.

In a linear modelling experiment, the value of $\chi^2$ should (locally) form a quadratic hyper-surface around the minimum. In such a case, the uncertainties on fitted parameters can be estimated directly from the properties of the surface [e.g. section 15.6 of \citet[][]{1992nrca.book.....P}].

An image generated through gravitational lensing is highly non-linear in the lens model parameters. Hence another technique such as Monte-Carlo simulations must be used to estimate the uncertainties on fitted parameters. For very small changes in the model parameters, however, the result might be `locally linear' so it is worthwhile examining the properties of the $\chi^2$ surface anyway.
If the $\chi^2$ surface locally approximates a quadratic, then quantifying a parameter's uncertainty is a much easier task than having to do extensive simulations.
In this section, a Monte-Carlo simulation is performed on the simulated optical lens to measure the true uncertainty in fitted parameters. These results are then compared to the values derived from the $\chi^2$ surface.

Using the same simulated optical image and noise as before, 400 datasets of the simulated optical source were generated.
For each simulation the best fitting parameters were found with the downhill simplex method \citep[e.g.][]{1992nrca.book.....P} using a SPEMD model with parameters ($b,q,\theta,\beta$) free to vary. The lens centre and core radius were fixed because the lens centre is usually well determined for an optical lens and there are no central images, so there is no point investigating a core.
A $15\times15$ source plane was used with default source pixel value (the $A$ parameter from \S \ref{sec:how_it_works}) fixed equal to the background noise level in the image.

\begin{table}
\centering
\begin{tabular}{l|rccc}
Parameter & True value & mean & std dev & Skewness \\ \hline
b (arcsec) & 1.2 & 1.200 &  0.0017 & 0.08 \\
$q$ & 0.82 & 0.80 & 0.02 & 1.73 \\
$\theta$ (degrees) & 68 & 68.03 & 0.42 & -0.19 \\
$\beta$ & 1.16 & 1.22 & 0.02 & 1.61 \\ \hline
$\chi^2$ &      & 1142 & 45 & 0.04
\end{tabular}
\caption{Statistics of the fitted parameters for 400 realisations of the simulated optical source.}
\label{tab:opt_mc_res}
\end{table}

\begin{figure*}
\centering
  \includegraphics[scale=0.22]{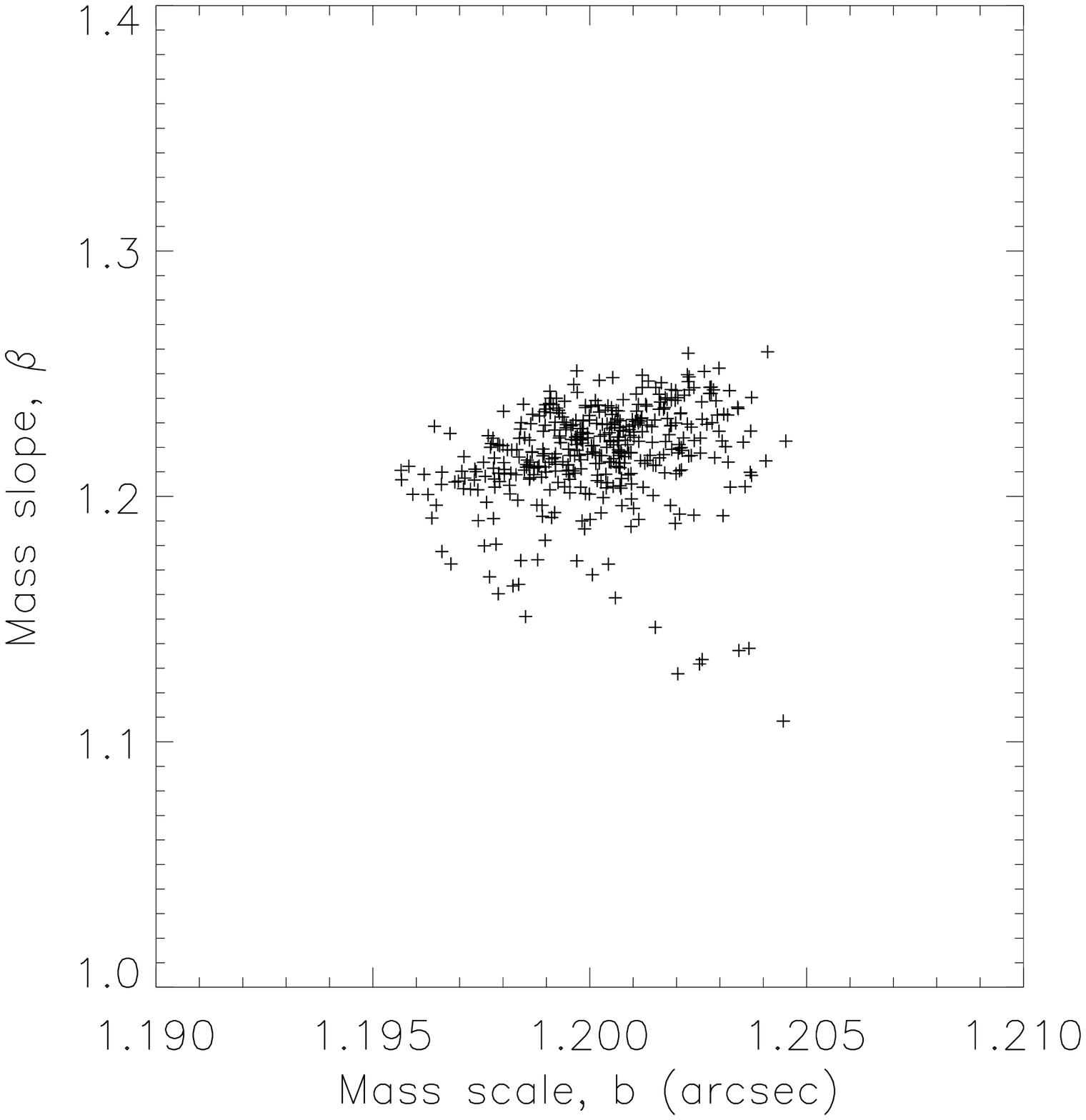}
  \includegraphics[scale=0.22]{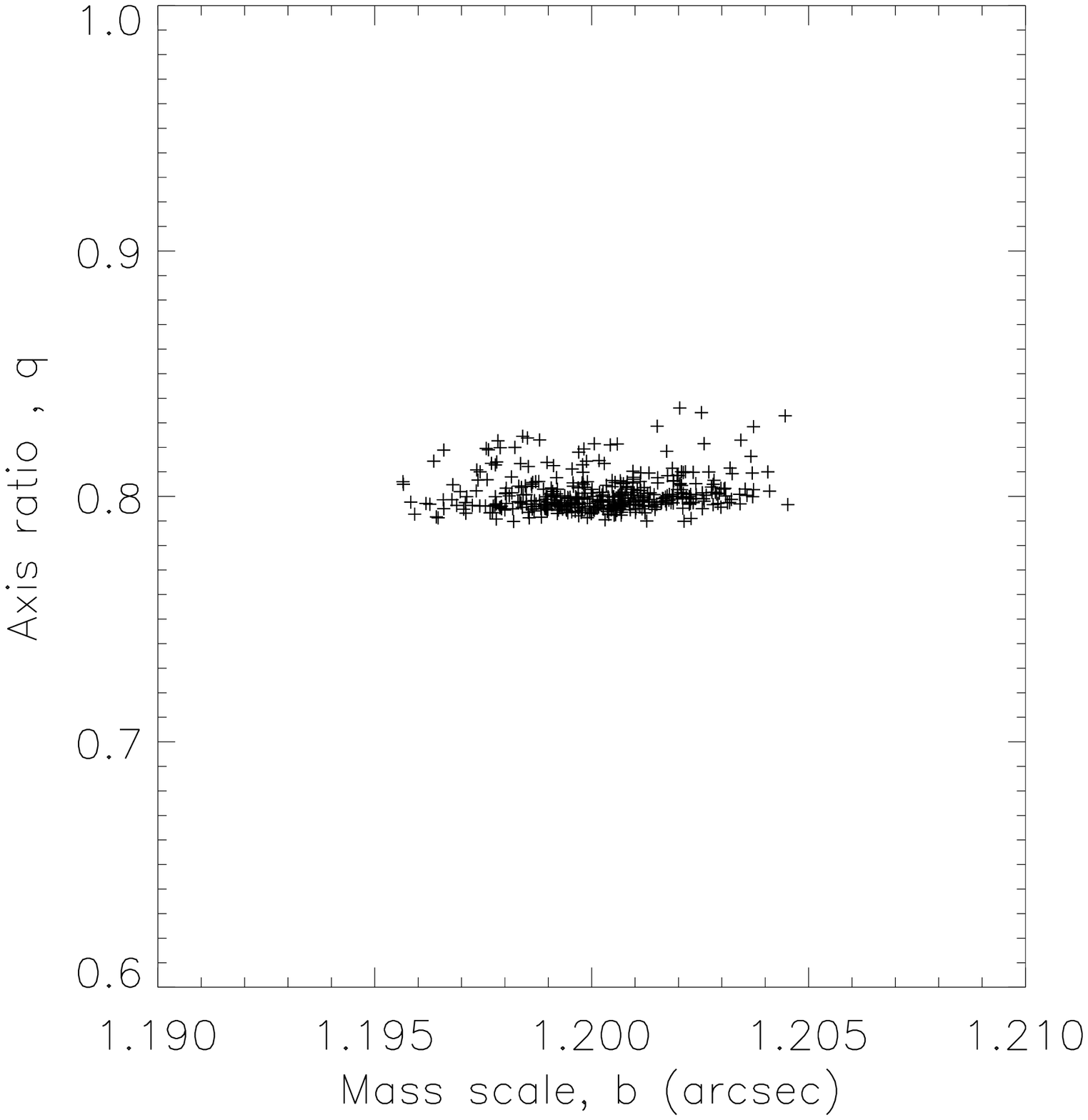}
  \includegraphics[scale=0.22]{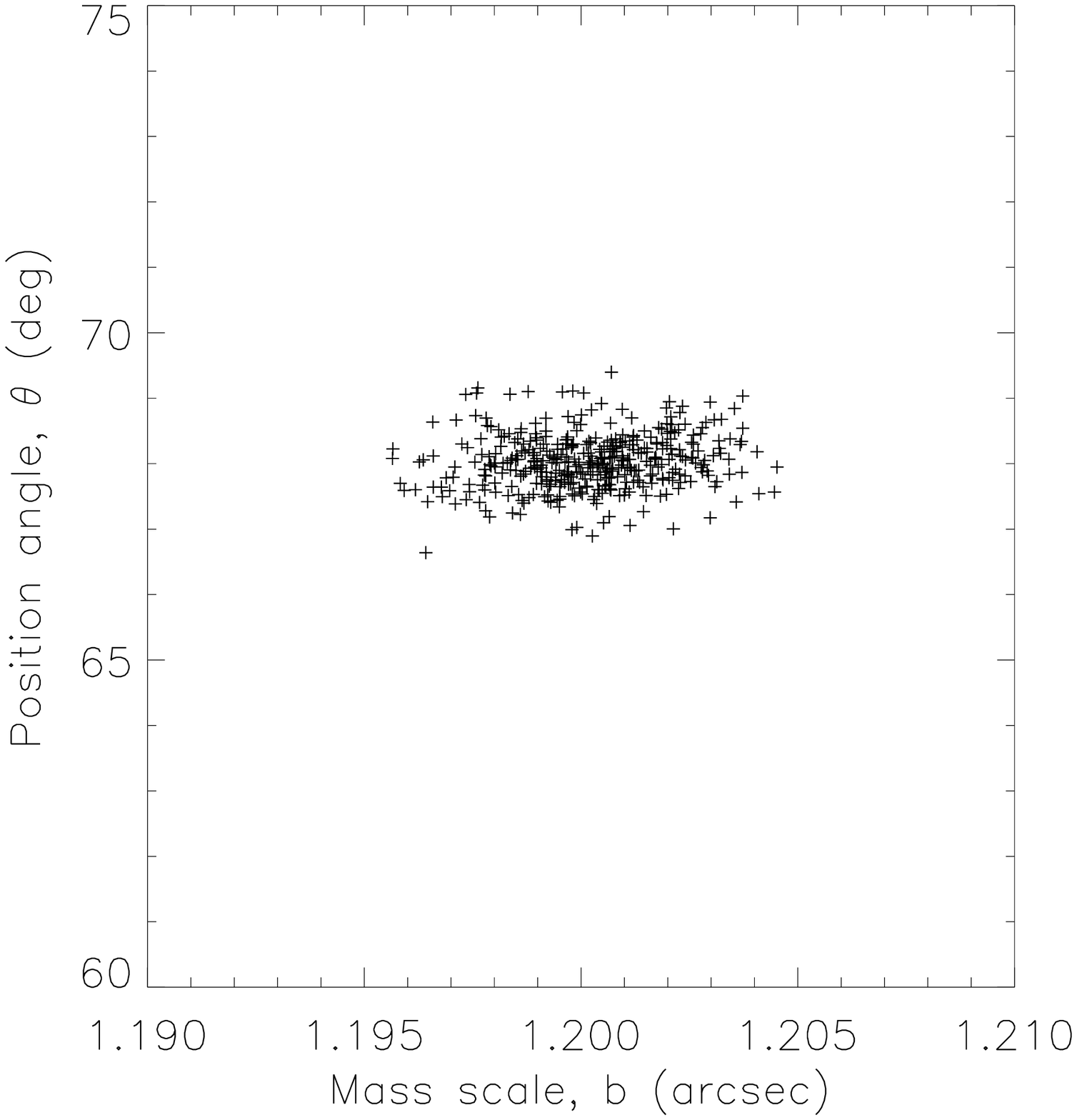}
  \includegraphics[scale=0.22]{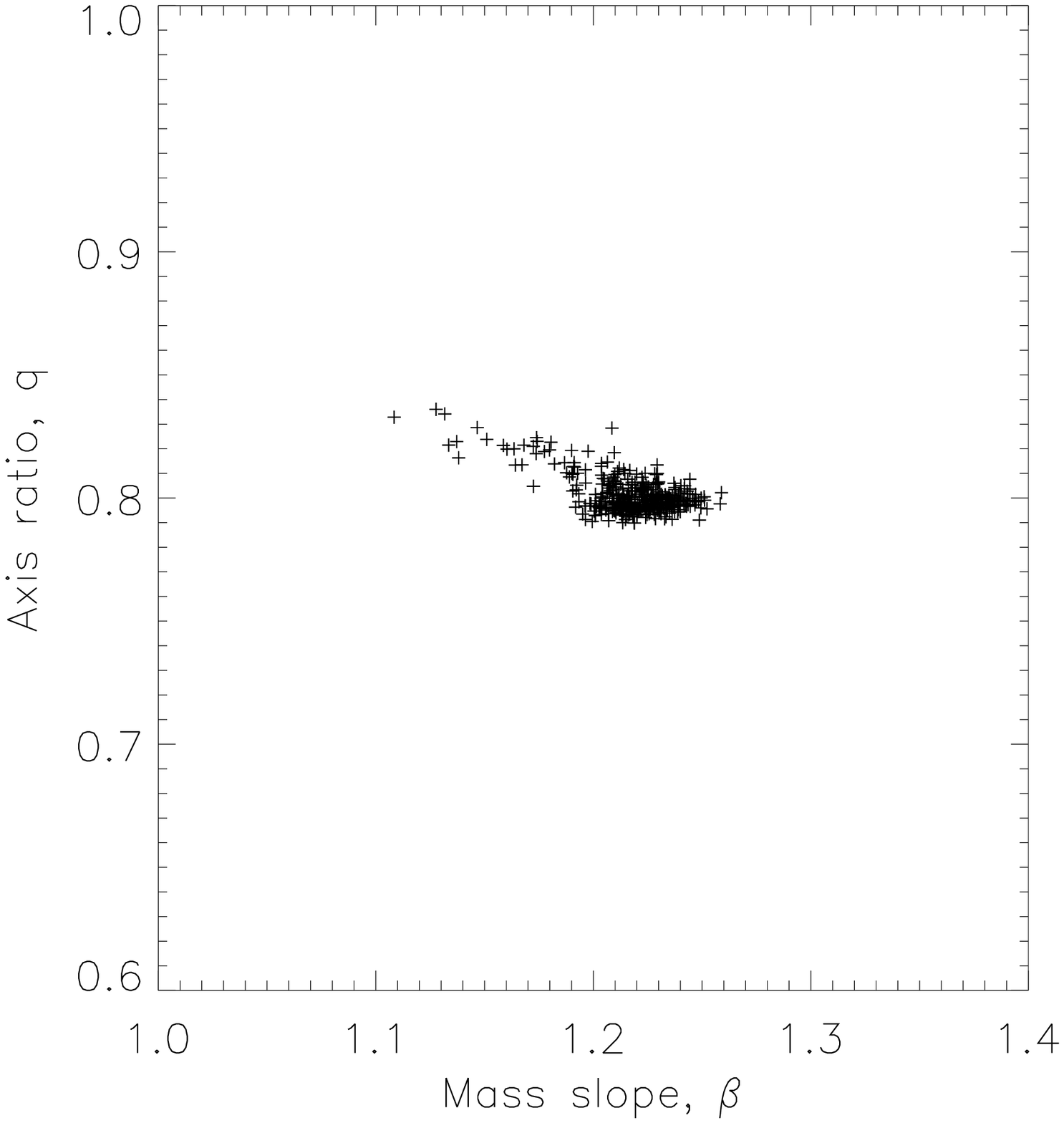}
\caption{Best-fit parameters over all the realisations of the simulated optical source.}
\label{fig:opt_mc_correlations}
\end{figure*}

The mean, standard deviation and skewness of the best-fit simulation parameters are shown in Table \ref{tab:opt_mc_res}.
The table shows that all of the lens model parameters have been well recovered although in the case of the axis ratio and mass slope, the best-fitting parameters are not distributed normally. The tail in the distribution increases the standard deviation (hence uncertainty) of a measurement. (A skewness of 1 makes the distribution approximately twice as wide compared to a normal distribution.)
Interpreting the standard deviation value as  `$1\sigma$' error, the mean of the best-fitting parameters are all within $1\sigma$ of the true value, with the exception of $\beta$. We note that of all the model parameters, only $\beta$ seemed to be affected by the starting position of the simplex search and even then the results only varied by $\beta \pm \sim$0.02.
Noteworthy is the correlation between $\beta$ and $q$ as revealed in the Fig. \ref{fig:opt_mc_correlations}. This correlation is related to the ellipticity/mass-power-law degeneracy found in position-only QSO lens models \citep{1994AJ....108.1156W}.

\begin{figure*}
\centering
  \subfigure[Marginalised contours of $\chi^2$]{
    \includegraphics[scale=0.25]{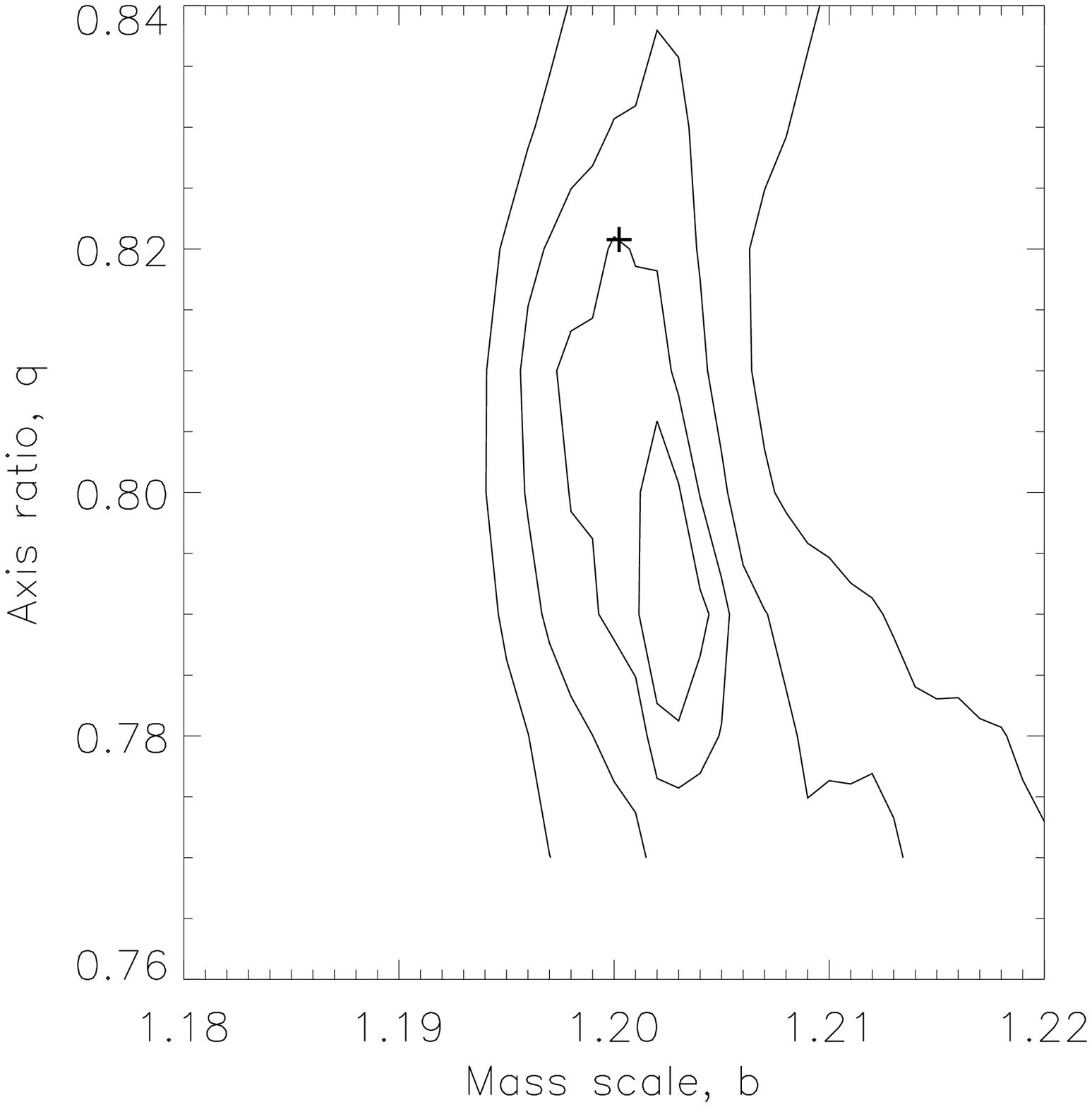}
    \includegraphics[scale=0.25]{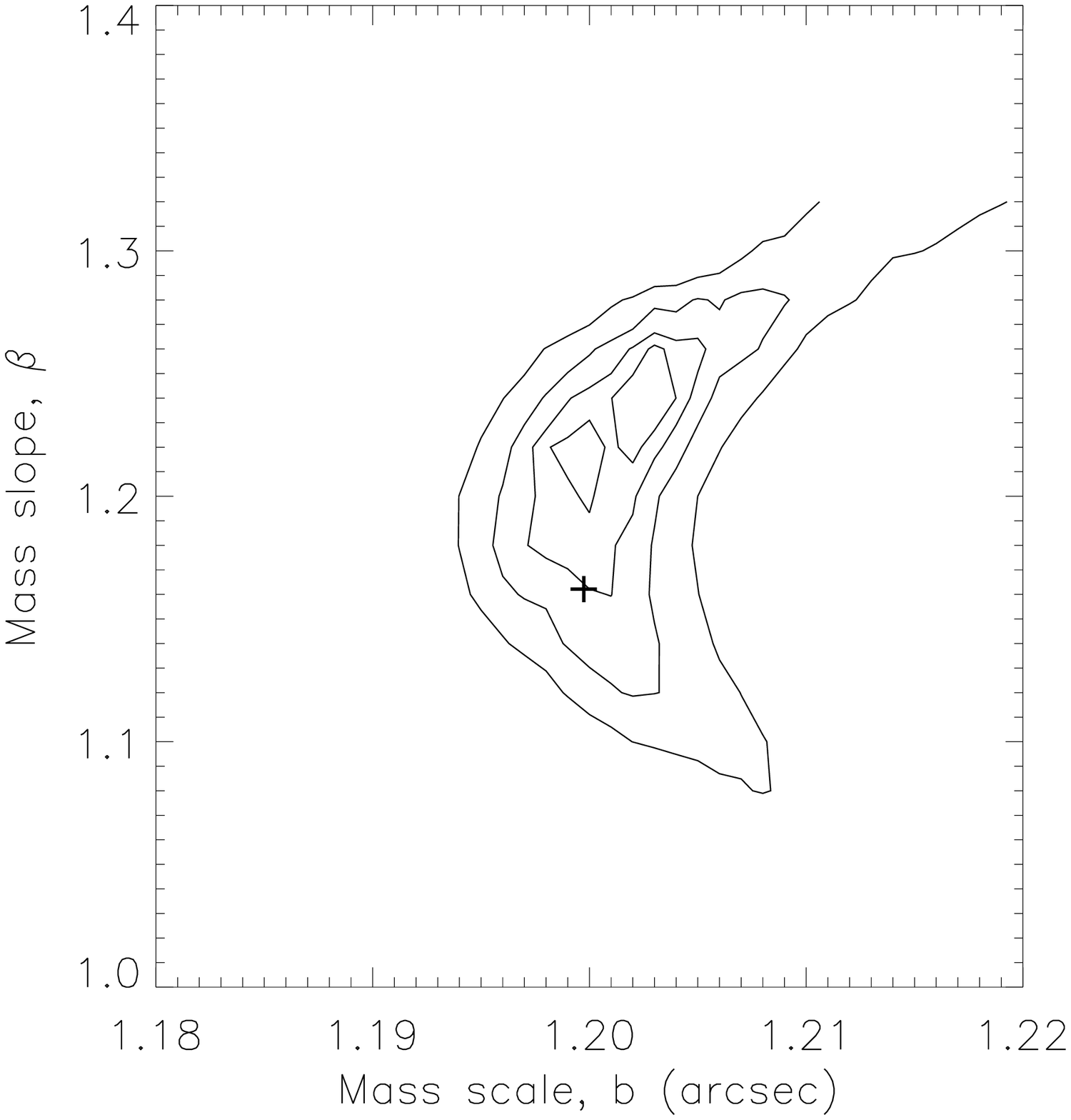}
    \includegraphics[scale=0.25]{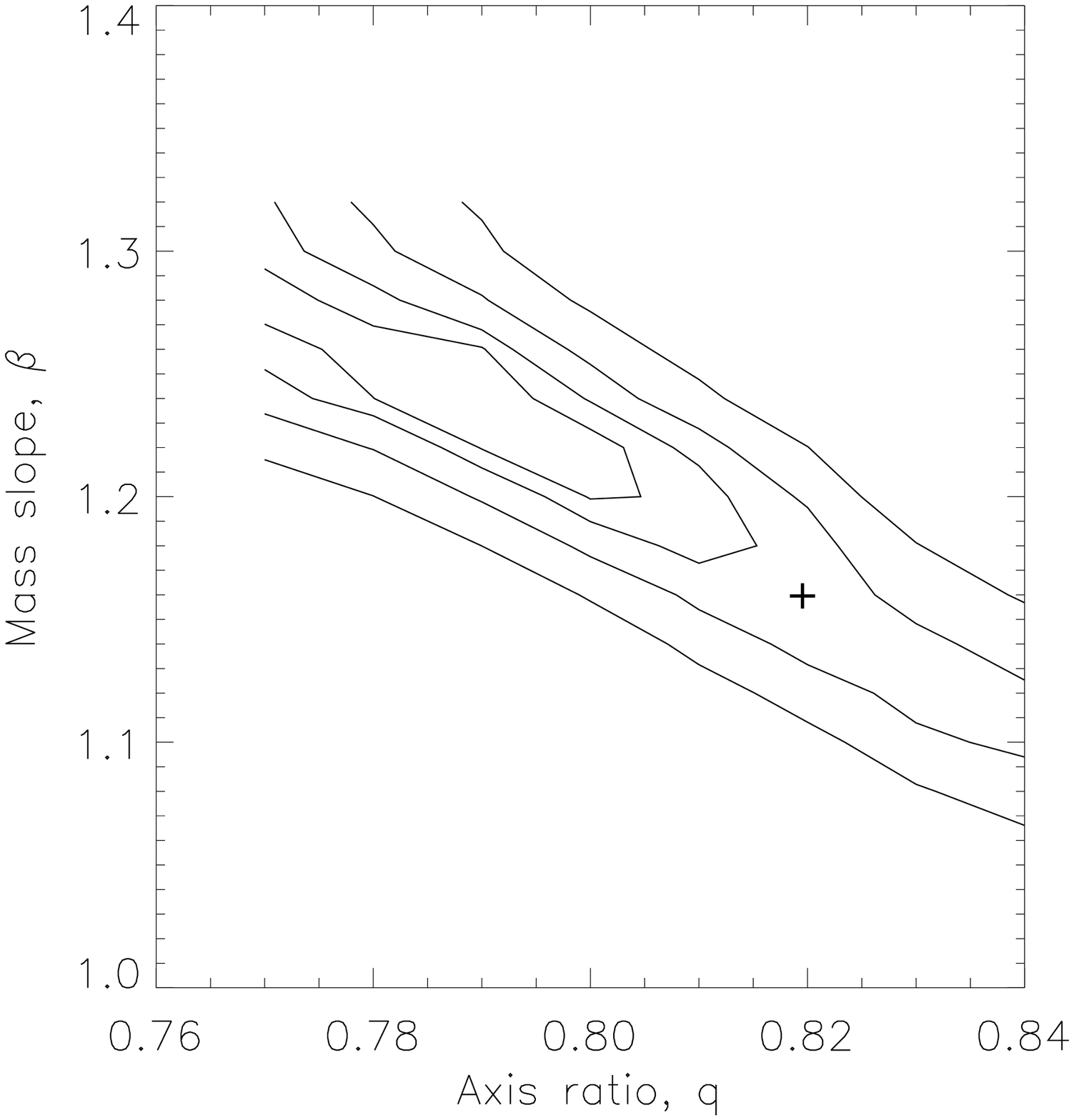}
\label{fig:opt_chimarg}
  } \quad
  \subfigure[Slices through $\chi^2$ space - not marginalised contours]{
  \includegraphics[scale=0.25]{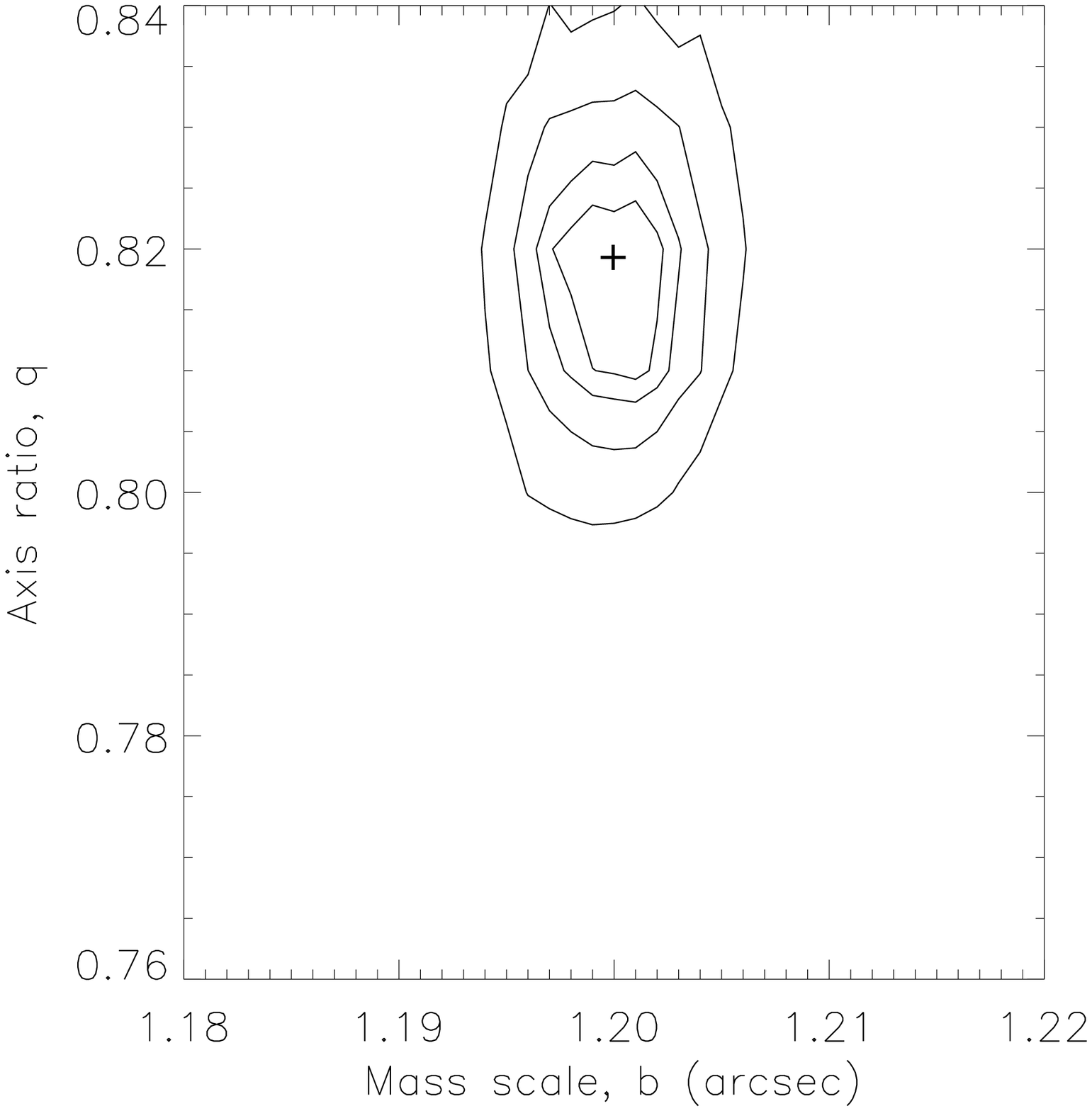}
  \includegraphics[scale=0.25]{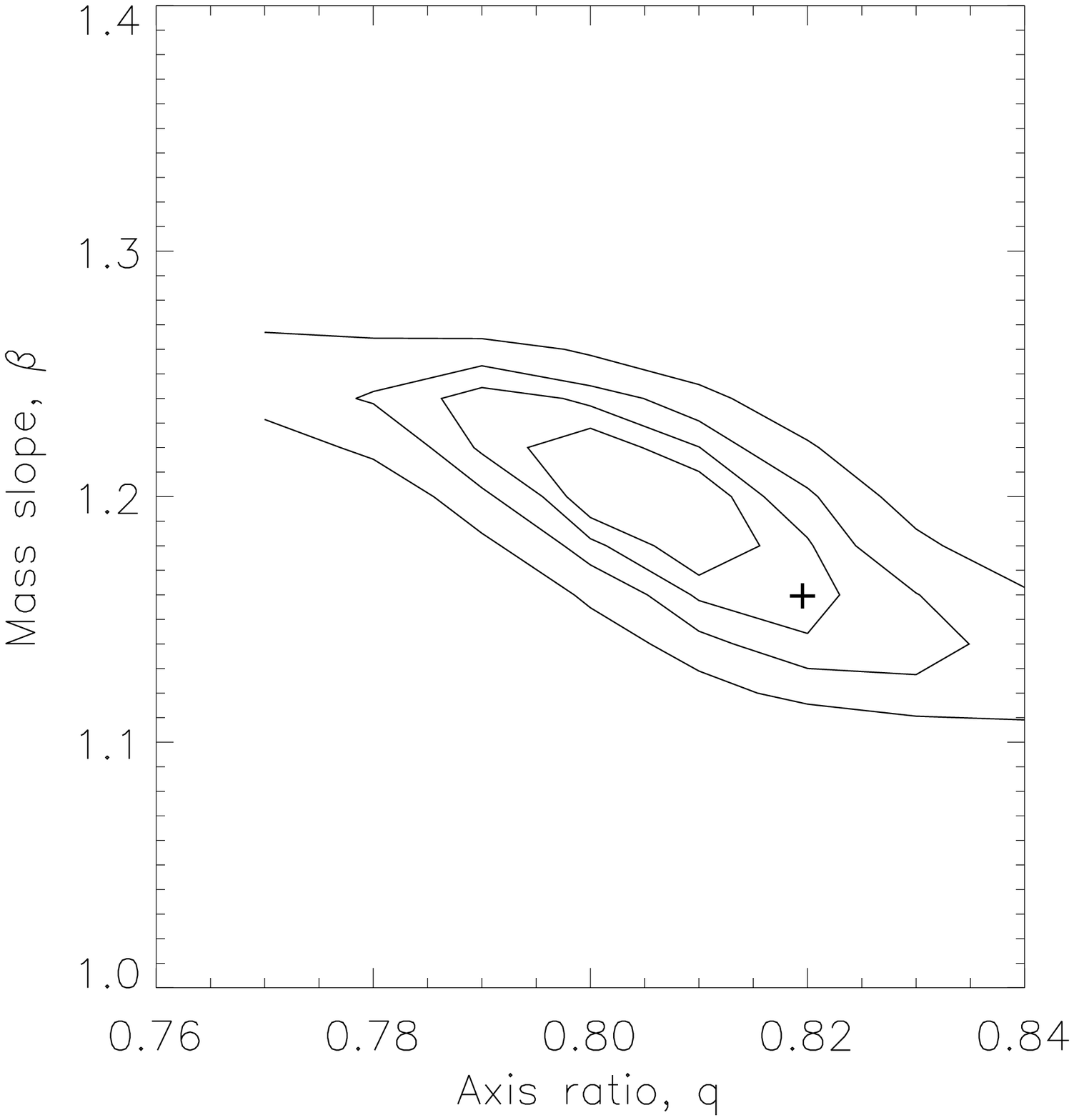}
\label{fig:opt_chislice}
  }
\caption{Contours of $\chi^2$ for different combinations of model parameters. Contours increase from the minimum by 2.3, 4.6, 9.2 and 18.4. Crosses show the parameter values used to create the data.}
\end{figure*}

Using a single simulation, parameter space was searched around the minimum to make a map of the $\chi^2$ surface.
Shown in Fig \ref{fig:opt_chimarg} are marginalised contours of $\chi^2$ for various combinations of parameters. The contours are 2.3, 4.6, 9.2 and 18.4 above the minimum which correspond to the 68\%, 90\%, 99\% and 99.99\% confidence intervals for the two parameters shown.

Examining the first two plots in Fig \ref{fig:opt_chimarg}, it is clear that the mass scale, $b$, is well constrained and that the 68\% confidence region (approx 1.200 to 1.204) corresponds well to an uncertainty of $\pm 0.002$ suggested by the Monte-Carlo simulations. The third figure of  Fig \ref{fig:opt_chimarg} shows a large uncertainty in $q$ and $\beta$- far more than is suggested by the Monte-Carlo simulations. In this case the $\beta-q$ degeneracy forms a valley through $\chi^2$ space, although it is reassuring that the simplex method used in the Monte-Carlo simulations has found its way down in most cases. Again, if we interpret the $1\sigma$ errors from the 68\% contour and from the Monte-Carlo simulations to be $\beta \pm 0.05$ then, the true value is less than $2\sigma$ from the minimum at $\beta=1.23, q=0.79$. The implication for measuring errors is that the long valley of 99.99\% uncertainty does not seem to affect the results. We conclude then, that the $\chi^2$ surface is a good measure for the statistical uncertainty of the lens model parameters.


As an illustration of the accuracy to which $b$ can be measured if $\beta$ is fixed, a slice (not marginalised) through the $\chi^2$ surface is shown in Fig. \ref{fig:opt_chislice} (LHS). In this case, the 68\% uncertainty is $0.004$ arcsec. Similarly, a slice through $\beta-q$ space (RHS) shows a well localised minimum. Hence, if extra constraints can be added to a lens model through (say) morphological or velocity dispersion information, the resulting lens model parameters will be more tightly constrained.


\subsubsection{Quantitative comparisons between lens models}

Because the properties of lensed images are determined mostly by the lens mass profile in the region of the images and the `correct' lens model has been used in these simulations, it is not surprising that the mass slope has been recovered well. This leads to the question: are there other lens models with a similar mass slope in the region of the images, but quite different elsewhere, that can fit the data? This question is the topic of this section.

While there are virtually limitless combinations of components that could comprise a mass model, the main models of interest here are the singular isothermal ellipsoid in external shear (SIE+$\gamma$) and the constant M/L model. These models are motivated by the fact that the image positions in almost all QSO lenses can be fitted with a SIE+$\gamma$ model \citep{1997ApJ...482..604K} and that a power-law mass distribution (SPEMD) has the potential to mimic the effects of a constant M/L profile under some circumstances (if the mass profile slope is locally constant at the image radius). The SIE is the same as a SPEMD with $\beta=1$. The external shear is parametrised by its magnitude, $\gamma$, and position angle, $\theta_{\gamma}$.

Testing the SIE+$\gamma$ model is important because external shears can be used to improve lens models which otherwise provide poor fits to the data.
If an external shear can mimic the effect of a non-isothermal mass distribution, then there is little hope that accurate mass profile slopes (i.e. $\beta$) can be measured using image data alone.

The case of the constant M/L model is much harder to test because there is no photometric data associated with the simulated optical source. Typically, the photometric properties of the lensing galaxy are quite well determined, hence a test for the acceptability of a constant M/L model is fairly straightforward: just build the lens model based on the photometric parameters and vary the M/L to find the best fit for an absolute goodness-of-fit measurement. (Whether the best fit M/L is realistic is another matter). For this test case, it will have to suffice to test whether a constant M/L model can mimic a power-law model rather than the other way around.

The constant M/L mass distribution is modelled with a S\'ersic \citep{1968adga.book.....S} profile. The  S\'ersic profile is defined by
\begin{equation}
\kappa(r) = \kappa_0 \exp^{-B(n) (r/r_s)^{1/n}}
\label{eq:sersic}
\end{equation}
where $\kappa_0$ is the central surface density in units of the lensing critical density ($\Sigma_{crit}=c^2 D_s / 4 \pi G D_d D_{ds}$), $r_s$ is the scale radius and $n$ is the shape parameter. The profile can also be elliptical with axis ratio $q$, and position angle $\theta$. The S\'ersic profile incorporates the de Vaucouleurs ($n=4$), exponential ($n=1$) and Gaussian ($n=1/2$) models with higher $n$ producing more cuspy central peaks. In this test, $n$ is fixed to 3 based on the photometric properties of the lensing galaxy in ER0047-2808 \citep{2005MNRAS.360.1333W} and for elliptical galaxies in general \citep{2003ApJ...582L..79B}.

\begin{figure}
\centering
\includegraphics[scale=0.35]{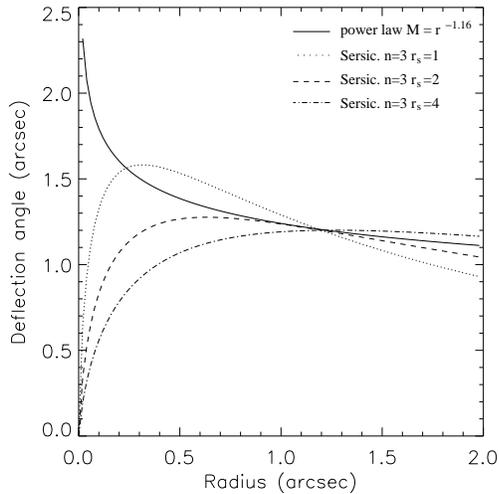}
\caption{Relative deflection magnitude for S\'ersic profiles with different shape parameters in comparison to a power law model.}
\label{fig:opt_defl_example}
\end{figure}

A constant M/L model that has the best chance of mimicking a power-law must have a mass slope which is locally constant and equal to $\beta$. Shown in Fig. \ref{fig:opt_defl_example} are plots of deflection angle vs radius for circularly symmetric S\'ersic profiles with $n=3$. Also shown is the deflection angle vs. radius of a power-law model with $\beta=1.16$. This figure shows that for $n=3$, S\'ersic scale lengths, $r_s$, between approximately 2 and 3 produce deflections that are very similar to the power-law model in the region of interest. Hence, our tests for the constant M/L model were limited to $n=3$ and a range of $r_s$ between 2 and 3.

Using the same simulated data as the previous section, parameter space was searched for SIE, SIE+$\gamma$ and constant M/L models. The results of the search are shown in Table \ref{tab:opt_altmass_res}.
\begin{table*}
\centering
\begin{tabular}{l|r|c}
Model & Best $\chi^2$ & Parameters \\ \hline
SIE & 1138 & $b=1.225$, $q=0.854$, $\theta=67.6$ \\
SIE + $\gamma$ & 1138* & $b=1.225$, $q=0.854$, $\theta=67.6$, $\gamma=0$ \\
Constant M/L & 1121 & $\kappa_0 = 57.9$, $q=0.85$, $\theta=67$ \\ \hline
SPEMD & 1107 & $b=1.200$, $q=0.809$, $\theta=67.7$, $\beta=1.16$ \small{(fixed)}
\end{tabular}
\caption{Best-fitting models for the simulated optical data with incorrect lens models. *Note: No models with improved $\chi^2$ could be found using an SIE with external shear.}
\label{tab:opt_altmass_res}
\end{table*}

The first result from this test is that adding an external shear to an SIE model cannot improve the fit compared to a plain SIE model. Hence, adding an external shear cannot help improve the model for this type of lensed image. This result is consistent with the fact that an external shear distorts the critical line away from being an ellipse. Since the image has flux at various points near the critical line, the image is sensitive to such distortions and can rule them out.

The second result from the test is that the constant M/L model can produce a result which is better than the SIE.
All of these models produce statistically acceptable results although the difference in $\chi^2$ reveals that the models are different with $> 99\%$ confidence. Hence, if this was real data the SPEMD model would be deemed the best model.

These tests have shown, therefore, that this type of optical image can indeed distinguish between lens models although the differences (in terms of $\chi^2$) are quite small. Statistically speaking, both the SIE and constant M/L models are acceptable in terms of the data alone.

\subsection{A test with simulated radio interferometer data}
\label{sec:test_radio}

In this section, a series of tests is performed with simulated radio data.
To make the tests realistic, the lobe must be `observed' with a simulated telescope then an image generated for the software to analyse. Typically, the image is generated with the CLEAN algorithm \citep{1974A&AS...15..417H,1980A&A....89..377C,1984AJ.....89.1076S,1984A&A...137..159S}, see also \citep{1999ASPC..180..151C}.
Differences between the real image and the CLEANed model can arise because of incomplete $(u,v)$ coverage and/or gain and phase errors introduced in the observing process. To distinguish between these effects, it is necessary to create simulated data for different length observations both with and without errors.

\begin{figure}
\centering
      \includegraphics[scale=0.3]{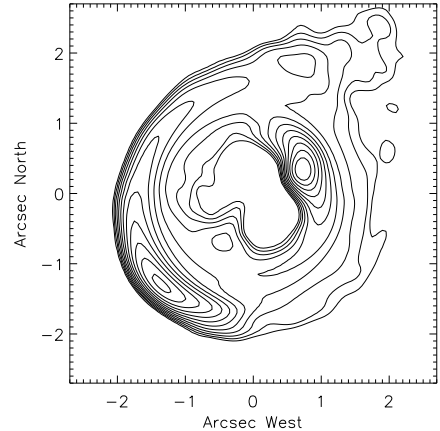}
\caption{The simulated lensed lobe `true' image. Contours increase by $\sqrt{2}$ from 0.1 mJy beam$^{-1}$.}
\label{fig:fake_lobe}
\end{figure}

Simulated radio data were created by taking the \emph{unlensed} radio lobe from MG1549+3047 (shown on the left in Fig. \ref{fig:recon_fakelobe}) as the `true' sky brightness (as seen by a noiseless telescope with infinite resolution). The lobe was corrected for any negative values then projected through a PIEP lens (defined by the lensing potential $\psi = b(r_c^2 + (1-\epsilon)x^2 + (1+\epsilon)y^2)^{1/2}$ with Einstein radius, $b$, ellipticity, $\epsilon$, orientation angle, $\theta$ and core-radius, $r_c$) with $b=1.35$\arcsec, $\epsilon=0.065$, $\theta=0^{\circ}$ and $r_c=0$.\footnote{For $r_c \ne 0$, $b$ is not the Einstein radius, it is simply a mass scale factor which is larger than the actual critical radius of the lens. Nevertheless, it will still be referred to as the `Einstein radius' or the `critical radius'.}
The location of the lobe (relative to the lens) was chosen such that the bright hot-spot was close to the inside edge of the radial caustic while the fainter, diffuse tail of the lobe passed near and through the centre of the lens. It was assigned a position in the sky near MG1549+3047. The simulated lensed lobe is shown in Fig. \ref{fig:fake_lobe}.

\begin{figure}
\centering
      \subfigure[1 hour integration]{\includegraphics[scale=0.3,angle=-90]{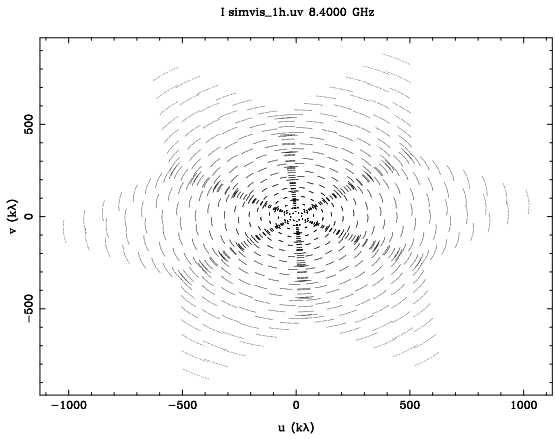}} \quad
      \subfigure[3 hour integration]{\includegraphics[scale=0.3,angle=-90]{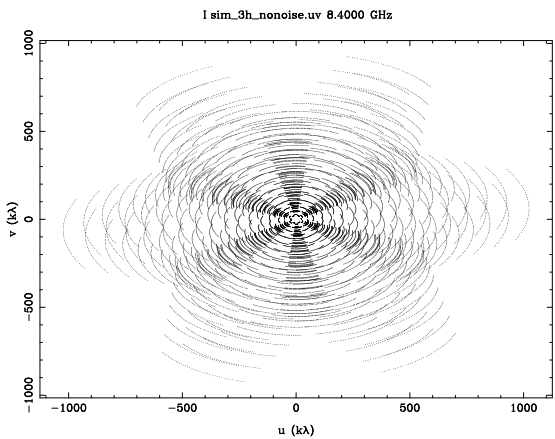}}
\caption[Simulated $(u,v)$ coverage]{$(u,v)$ coverage for simulated 1 hour (top) and 3 hour (bottom) observations of the lensed lobe.}
\label{fig:fake_obs_uv_coverage}
\end{figure}

Using the \textsc{Miriad} tasks `uvgen' and `uvmodel', visibilities were generated for observations of the simulated lobe with the VLA A array at 8.4 GHz.
Observations were generated both with and without realistic gain and phase noise for 1-hour and 3-hour integrations. Observation time was spread approximately evenly around HA = 0.
Plots of the $(u,v)$ coverage for both integration times are shown in Fig. \ref{fig:fake_obs_uv_coverage}.

\subsubsection{Image reconstruction with ideal and realistic data}
Images from interferometers with incomplete $(u,v)$ coverage must be reconstructed with some algorithm like CLEAN. This introduces known artifacts into the image [see \citet{1999ASPC..180..151C} for a detailed discussion]. In the context of lens modelling, it is the effects of these artifacts on the lens model that are of interest.

\begin{figure}
\centering
      \includegraphics[scale=0.3]{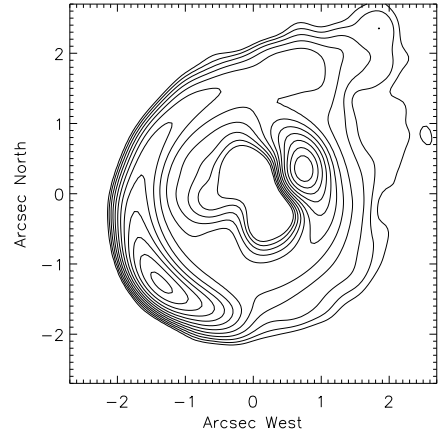}
      \includegraphics[scale=0.3]{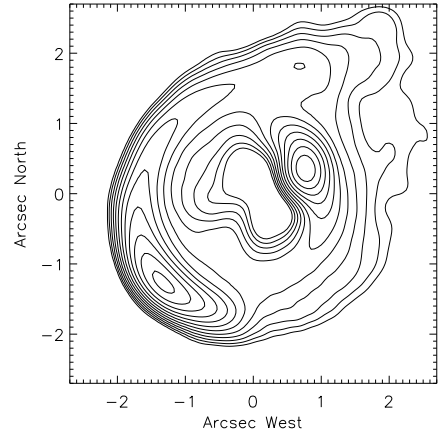}
\caption{Top: The simulated lensed lobe convolved with a Gaussian restoring beam which is the same size as that generated from the 1 hour simulated observation. This is what a perfectly CLEANed image should look like. Bottom: The CLEANed image from a 3 hour simulated observation of the lensed lobe with no observation errors. Contours increase by $\sqrt{2}$ from 0.1 mJy beam$^{-1}$.}
\label{fig:cleaned_lobes1}
\end{figure}

\begin{figure}
\centering
      \includegraphics[scale=0.3]{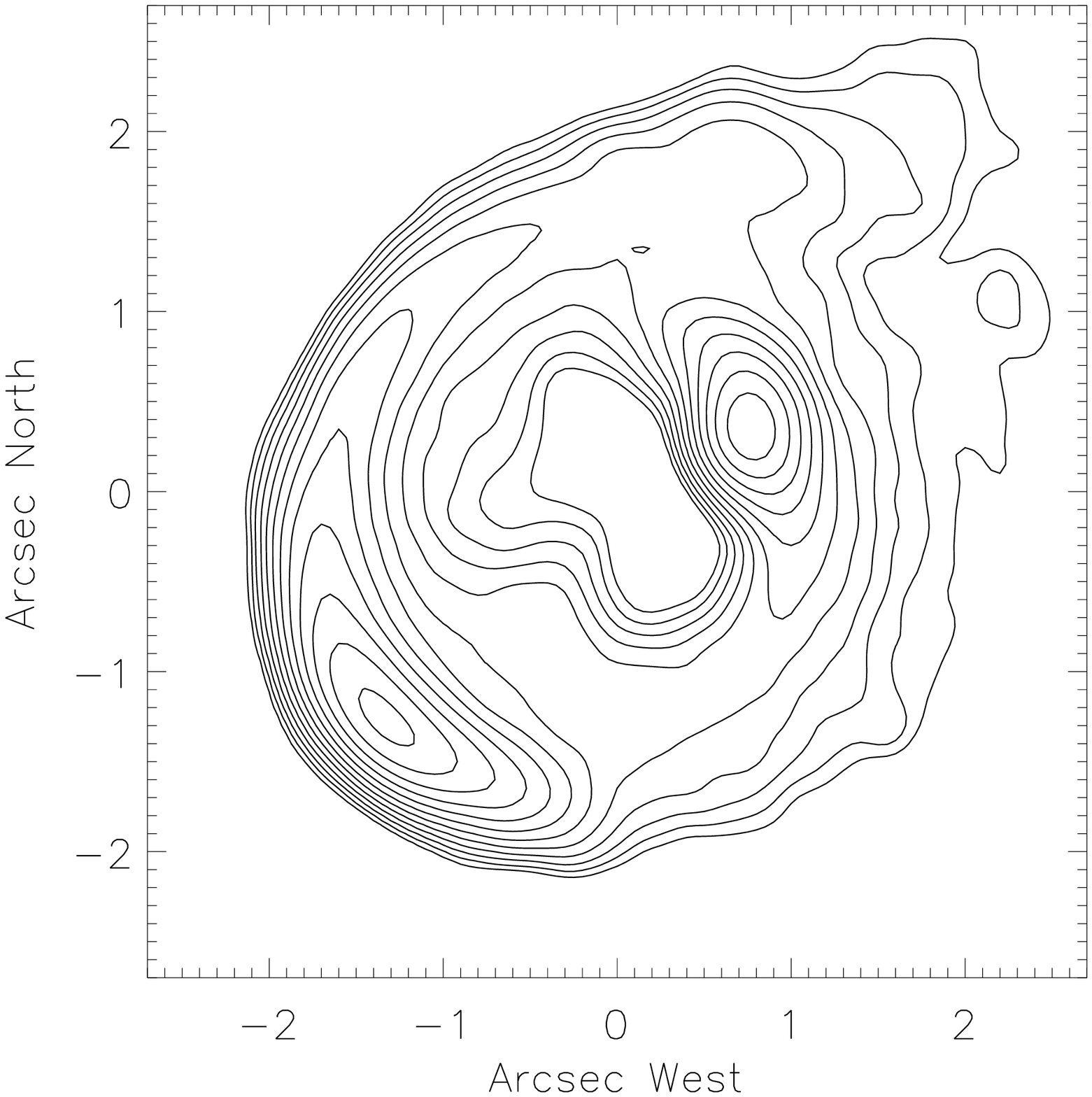}
      \includegraphics[scale=0.3]{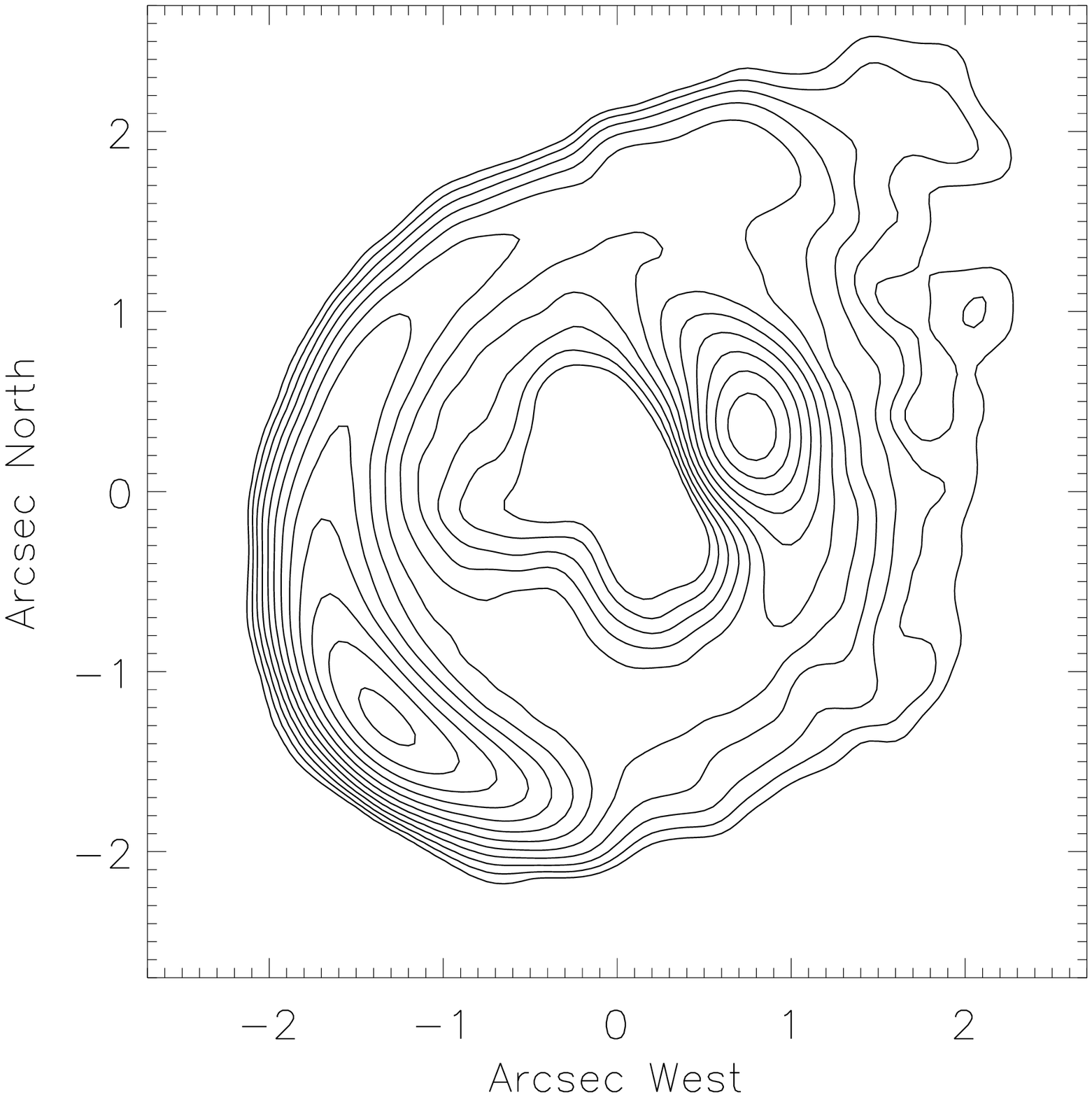}
\caption{CLEAN reconstructions of the simulated observations of the lensed lobe. Top: The image from a 1 hour simulated observation with no errors. Bottom: The image from a 1 hour simulated observation with gain/phase errors.  Contours increase by $\sqrt{2}$ from 0.1 mJy beam$^{-1}$.}
\label{fig:cleaned_lobes2}
\end{figure}

 The images were processed with the \textsc{Miriad} tasks `invert' `clean', `selfcal' and `restor' using 10000 iterations, natural weighting, only positive clean components and a small (0.02) loop gain. The field size was $512\times512$ pixels of size 0.05 arcsec and the beam FWHM $0.28 \times 0.27$ arcsec.
For the simulated observations that contained gain/phase errors, the `selfcal' task was used to correct the complex gains using self-calibration.
Calibration solutions for the VLA are over-constrained, therefore well determined.
In all cases of self calibration, the model converged in a single iteration.
Shown in Figs. \ref{fig:cleaned_lobes1} and \ref{fig:cleaned_lobes2} are the CLEANed images for the simulated observations along with the `true' image convolved with a Gaussian beam of the same size as the CLEAN restoring beam (i.e. the ideal reconstruction).

\begin{figure}
\centering
      \includegraphics[scale=0.7]{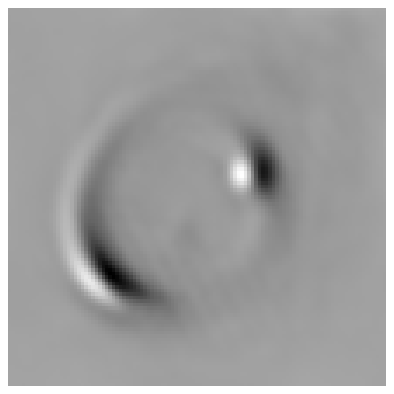}
      \includegraphics[scale=0.7]{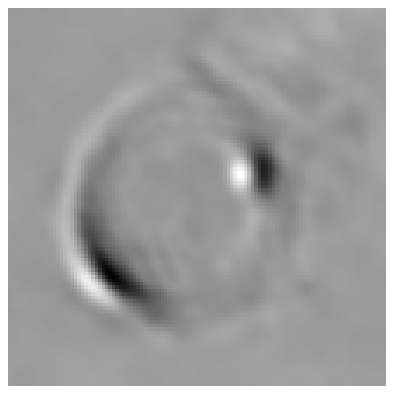}
      \includegraphics[scale=0.7]{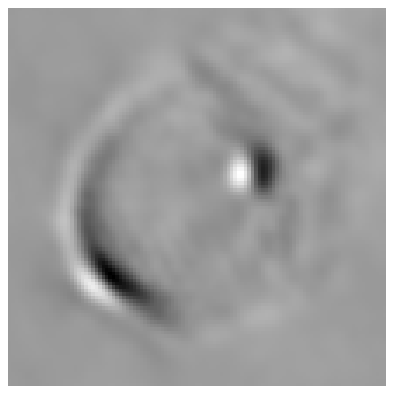}
\caption{Differences between the beam-convolved true lens image and the CLEAN reconstructions. Left: the 3 hour observation with no errors. Centre: The 1 hour observation with no errors. Right: A 1 hour image with gain/phase errors. In each case, the peak of the difference is approximately 10\% of the peak in the image, or $\sim 20$ mJy.}
\label{fig:model_vs_clean_diff}
\end{figure}

On the contour plots, the differences between the images are most obvious in the region of fainter, diffuse emission north and west of the inner image. However, it is not the faint, low magnification emission that provides the constraints for the lens model, it is the bright arcs and ring.
In Fig. \ref{fig:model_vs_clean_diff} the differences [i.e. the ideal image of Fig. \ref{fig:cleaned_lobes1} (top) minus the reconstructed image] between the beam-convolved (`true') image and the CLEANed images are shown as images, not contours. The artifacts generated by CLEAN are more apparent in these images. The large arc has a thin shadow around most of the bright region and the inner bright image is misshapen. Because the maximum/minimum of the difference between the bright regions in the images is on the same side for both images, the effect is that the image appears to be shifted. However, this effect is a coincidence. There has been no translation of the CLEANed image. The effect is likely a consequence of the finite pixelised grids generated by Fourier transforming and CLEANing the data. CLEAN components are chosen from a grid, so there must always be a small (fraction of a pixel) difference in position between the peak flux suggested by the data and the location of the corresponding CLEAN component.
It is apparent from these images that the main source of error in the image reconstruction comes from incomplete $(u,v)$ coverage and CLEANing, not observational/instrumental errors.

Given that most of the information about the lens model will come from the bright regions, the outcome of the fitting process is likely to be affected. In particular the apparent translation of the reconstructed images will certainly affect the determination of the lens centre. It is also worth noting that the self-calibration process can lose absolute astrometric accuracy in radio data so radio images should be aligned with optical whenever possible to avoid this problem.

\subsubsection{The effects of CLEANing on lens modelling}

To quantify the effects on the lens model of the artifacts discussed in the previous section, a series of Monte-Carlo simulations was performed. One hundred simulated observations of the simulated lensed lobe from the VLA A array at 8.4GHz were generated with realistic system noise, gain/phase errors and an integration time of 1 hour.
The observations were processed by an automated script using the same series of steps as described in the previous section, including self-calibration.

The lens was modelled as a PIEP with all parameters free to change.
Given the artifacts generated in the CLEANed images, in particular the apparent offset of the reconstructed image compared with the true image, the simulations were performed twice: once with the lens centre free to change and once with the lens centre fixed.
The PSF was taken directly from the Gaussian restoring beam and the noise in each pixel was set to the fixed theoretical RMS noise of $7.5\mu$Jy beam$^{-1}$.\footnote{It should be noted that the uncertainty in pixels in a CLEANed image is not well determined and the theoretical RMS noise is certainly a lower limit to the true noise in the pixels --- especially in the brighter regions. Nevertheless, without a solid theoretical understanding of the noise in CLEANed images only ad-hoc schemes can be used to model pixel noise in image space. Hence, using the theoretical noise is a good starting point. An alternate value is the off-source RMS noise in the CLEANed map, which is approximately $55\mu$Jy beam$^{-1}$ for the 1 hour simulations.}
The best-fitting parameters for the lens model was found using the downhill simplex method \citep[e.g.][ chapter 10]{1992nrca.book.....P}.

Before discussing the results of the simulations, there is an important question that must be addressed: how do we know we have the best model? In other words, is there a well defined minimum in $\chi^2$ which is defined as the best model? To explore this question, an initial investigation was undertaken with a single simulated image.
Parameter space was searched to understand the properties of the $\chi^2$ surface as a function of parameter value.

\begin{figure*}
\mbox{
	\subfigure{\psfig{figure=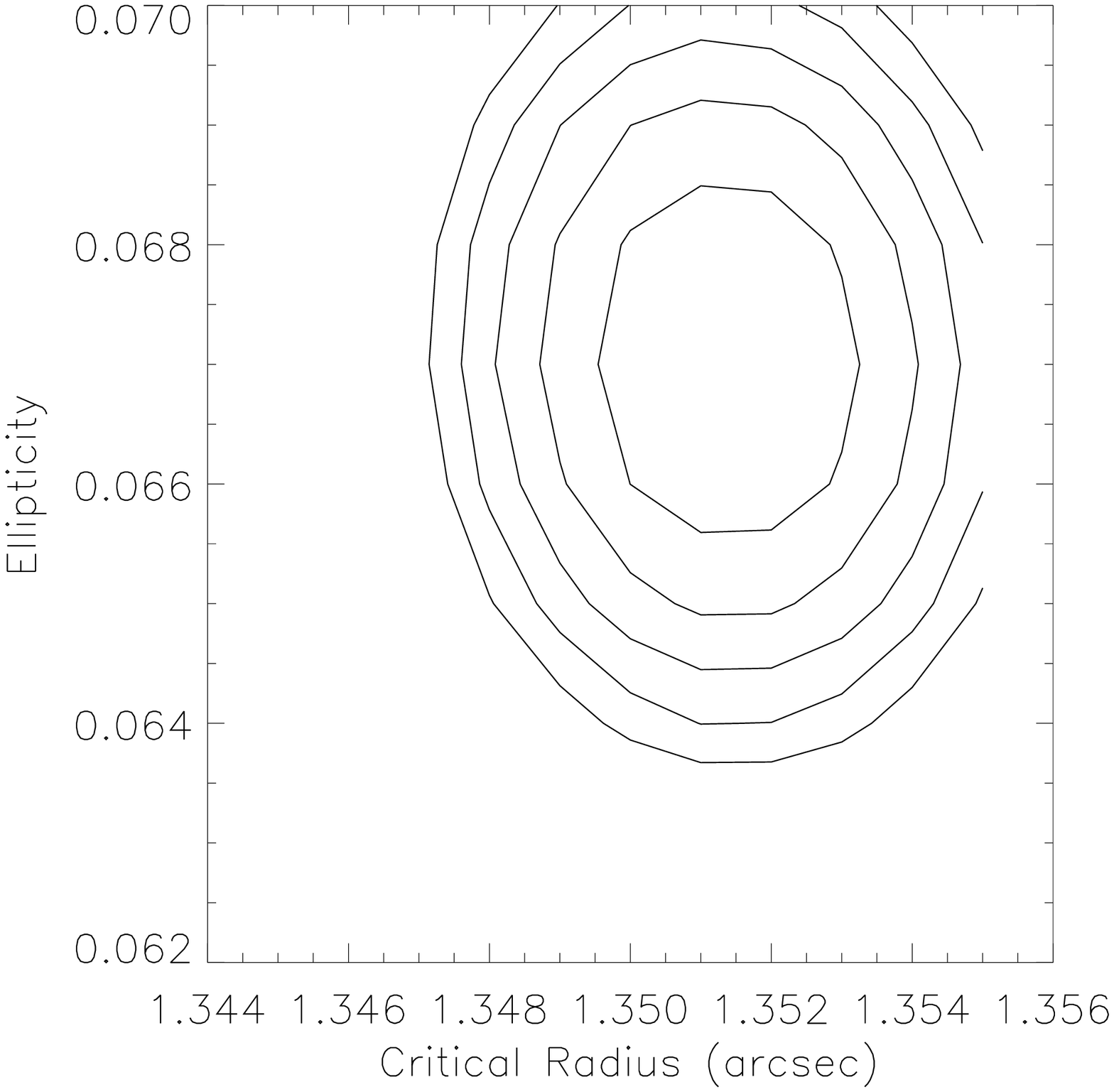,width=62mm}}
	\subfigure{\psfig{figure=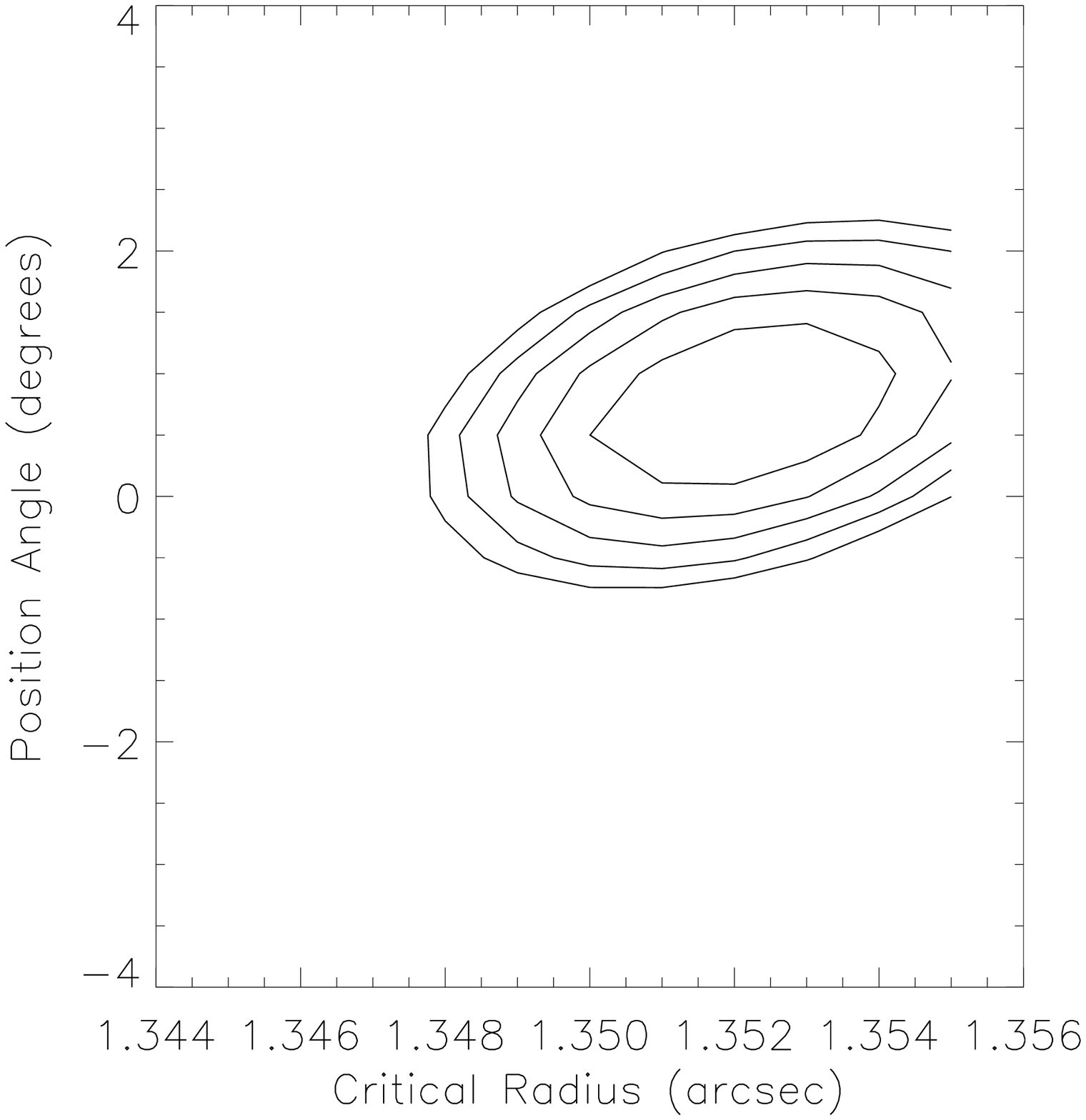,width=62mm}}
}
\caption{Cuts through the $\chi^2$ surface of the simulated lensed lobe. Parameters which do not vary are fixed to their best-fit value. Contours increase by 5\%, 10\%, 15\%, 20\% and 25\% from the minimum. Left: Critical radius (b) vs ellipticity ($\epsilon$). Right: critical radius (b) vs position angle ($\theta$).}
\label{fig:chisurf_kap1}
\end{figure*}

\begin{figure*}
\mbox{
  \subfigure{\psfig{figure=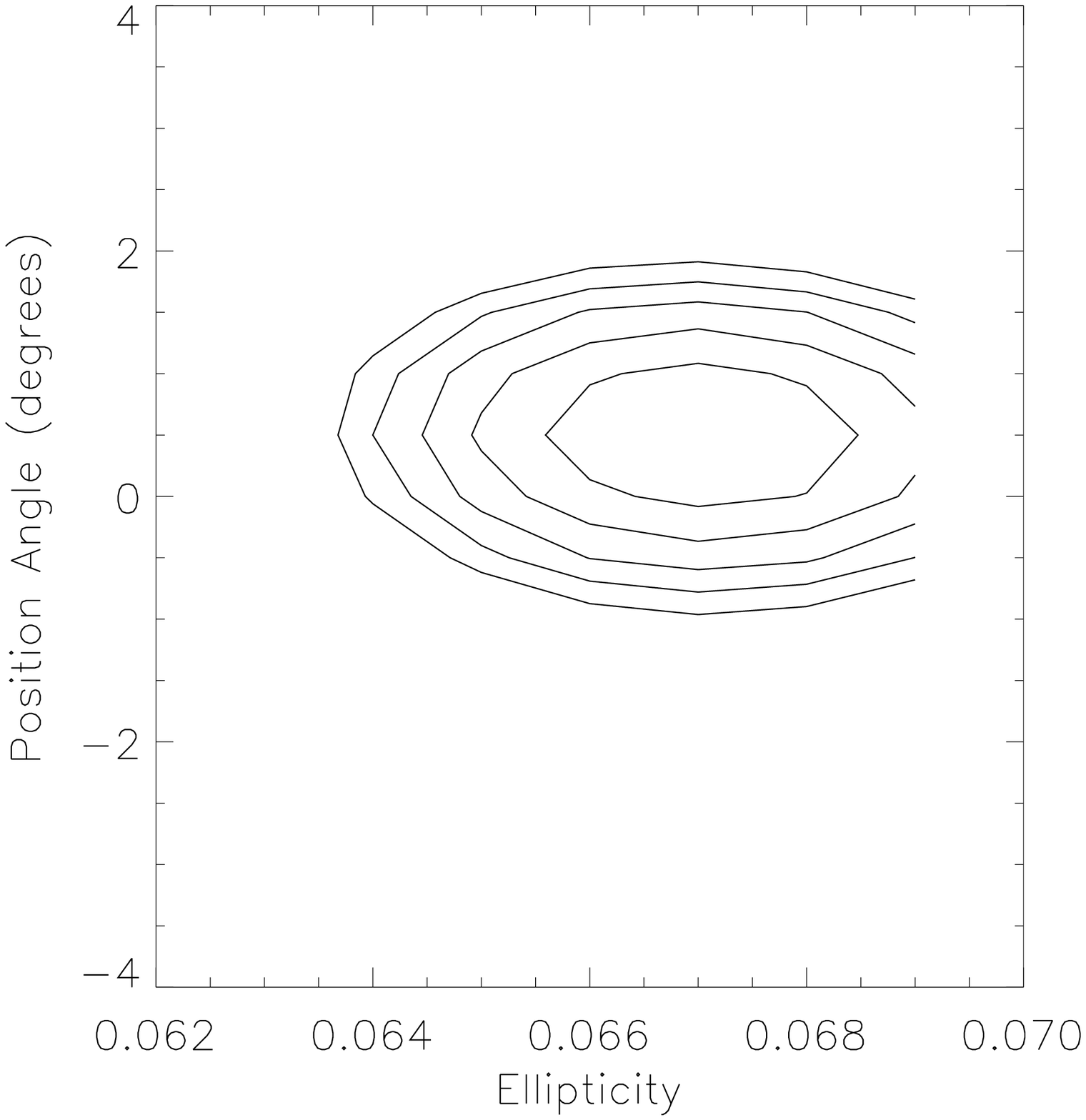,width=60mm}}
	\subfigure{\psfig{figure=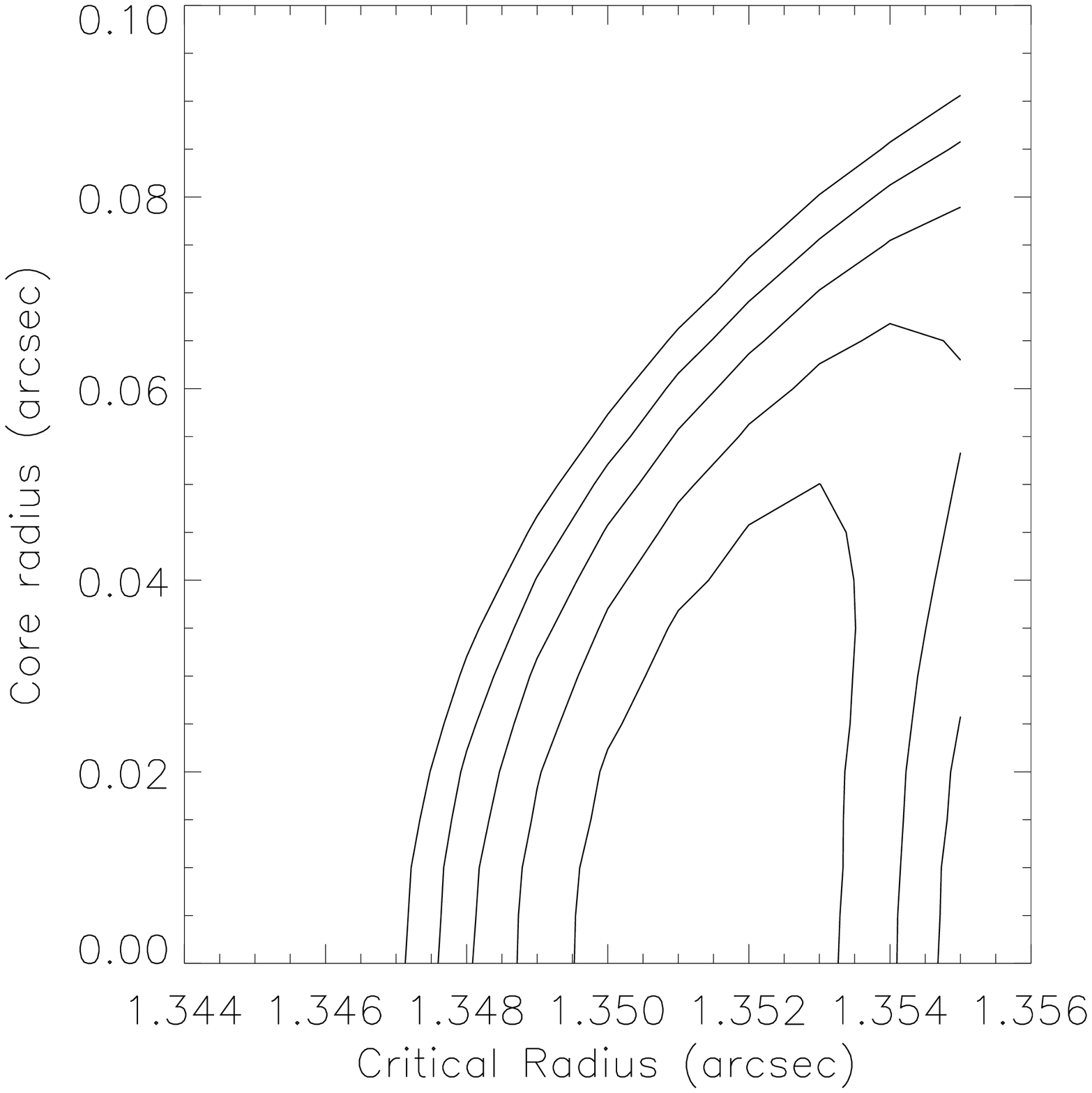,width=60mm}}
}
\caption{Cuts through the $\chi^2$ surface. Contours increase by 5\%, 10\%, 15\%, 20\% and 25\% from the minimum. Left: position angle ($\theta$)  vs ellipticity ($\epsilon$). Right: Critical radius (b) vs core radius ($r_c$).}
\label{fig:chisurf_kap2}
\end{figure*}

\begin{figure*}
\mbox{
      \subfigure{\psfig{figure=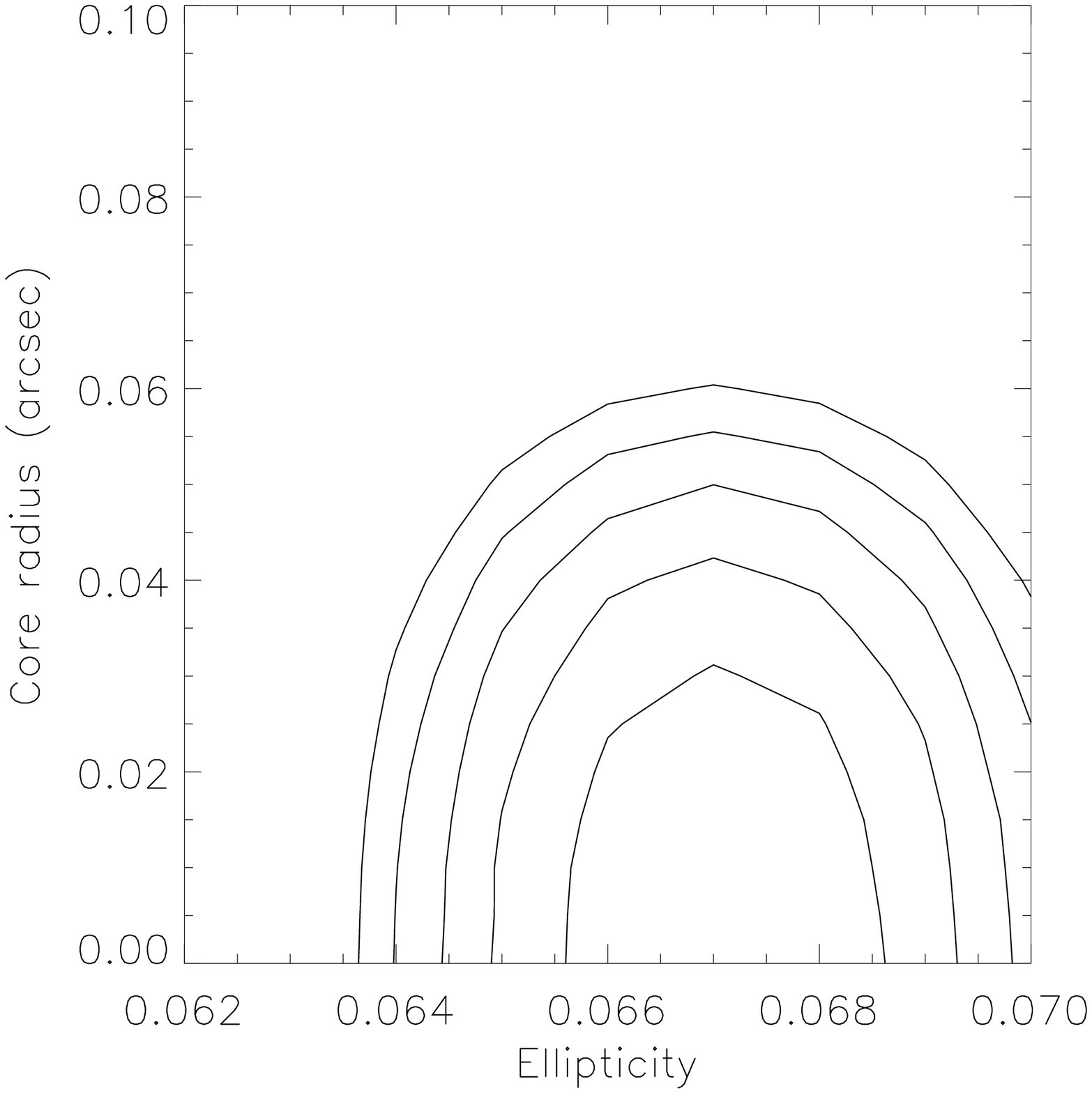,width=60mm}} 
      \subfigure{\psfig{figure=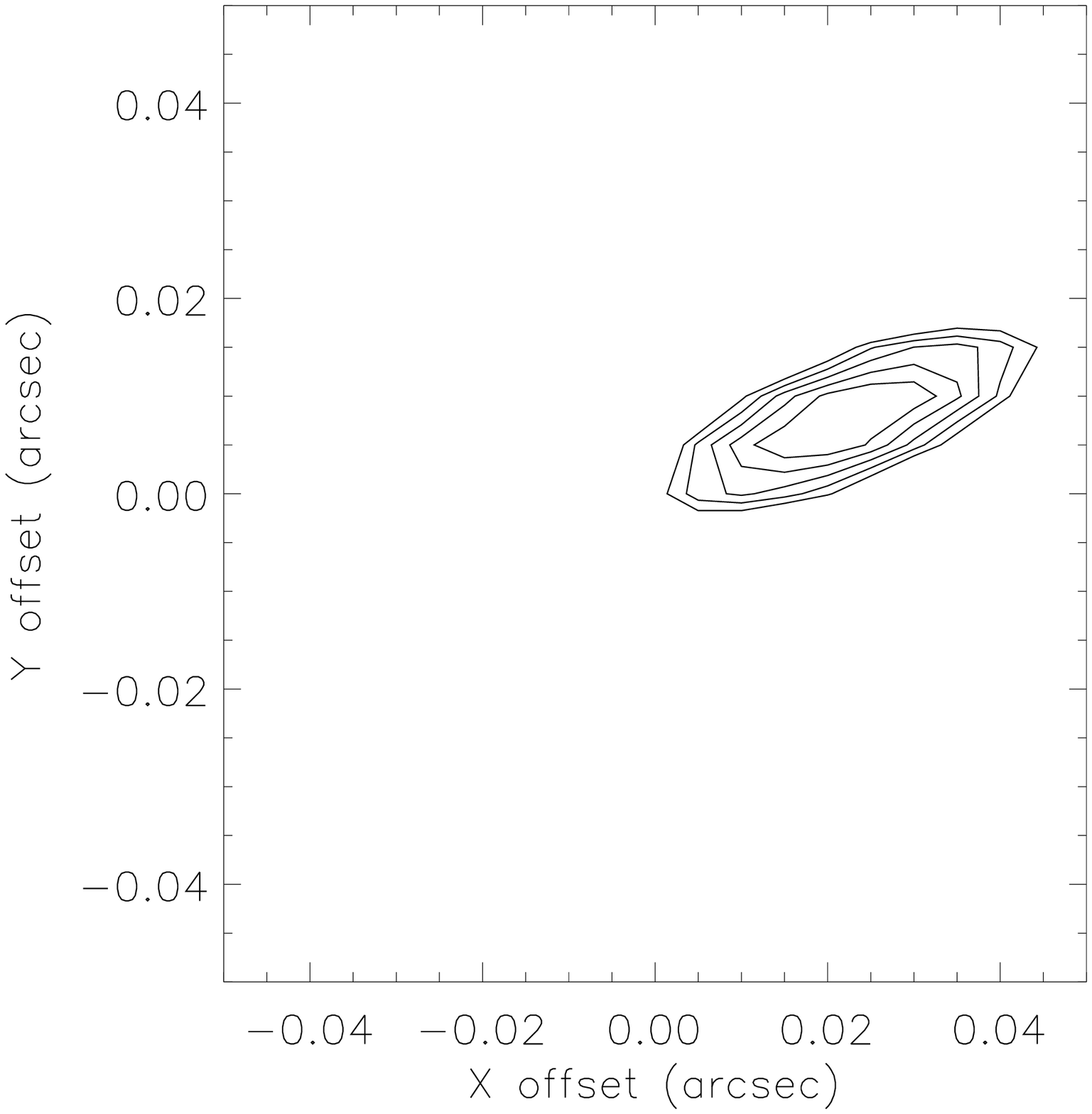,width=60mm}
}
}
\caption{More cuts through the $\chi^2$ surface of the simulated lensed lobe.}
\label{fig:chisurf3}
\end{figure*}

Figs. \ref{fig:chisurf_kap1}, \ref{fig:chisurf_kap2} and \ref{fig:chisurf3} show slices through $\chi^2$ space for the critical radius, ellipticity, orientation angle, core radius and lens centre.
In each case, the surface varies smoothly with parameter value and has a well defined minimum.
The best-fitting parameters are thus defined unambiguously by the minimum in the $\chi^2$ surface\footnote{One should always bear in mind the possibility of degeneracies between more than 2 parameters. In this case, a mass slope parameter is not being used so marginalisation over this parameter is not necessary}.
In addition, the smooth nature of the surface and well defined minimum means we can be confident that the search algorithm will find the global minimum in $\chi^2$.

It is apparent from Figs. \ref{fig:chisurf_kap1}, \ref{fig:chisurf_kap2} and \ref{fig:chisurf3} that there are some mild degeneracies between fitted parameters. There is a degeneracy between critical radius and core radius, and weak degeneracies between critical radius vs ellipticity and critical radius vs position angle.
The degeneracy between critical radius and core radius is to be expected because
a non-singular core can only be distinguished if it produces central images or changes the local slope of the mass profile enough to distort the inner image. In the case of the simulated lensed lobe, the bright inner image is $\sim 0.7$ arcsec from the centre, hence it is reasonable that the models produce the same result for small ($< 0.1$ arcsec) core radii. The $\chi^2$ surface in Fig. \ref{fig:chisurf_kap2} slopes away to the right because the overall mass scale (i.e. $b$) must be increased to compensate for the mass `lost' in the centre of the lens due to the non-singular core. However, the actual critical radius and total mass within the critical radius remain the same.
In the case of critical radius vs orientation angle, the effect is very small and probably due to the particular geometry of this simulated lens (i.e. it is essentially a double). If the lens was a quad or contained more bright regions it is difficult to see how the position angle could compensate for changes in the critical radius.

For the case of critical radius vs ellipticity, again this is a small effect. A small change in critical radius can be compensated for by a change in ellipticity for two-image systems. This moves the critical line which can move images relative to the critical line. However, a ring image such as that of the simulated lobe contains a great deal of information about the ellipticity. Hence it is well constrained as seen in Fig \ref{fig:chisurf_kap1}.

For the lens centre, the major axis of the $\chi^2$ contours are aligned with the angle between the centre of the source plane and the bright regions in the image. The effect on the images of moving the lens centre can be partially corrected by moving the source.

\begin{table}
\centering
\begin{tabular}{l|l|ll|ll}
 & True &\multicolumn{2}{c}{Lens centre free} & \multicolumn{2}{|c}{Lens centre fixed} \\
  Parameter  & value & Mean & std dev & Mean & std dev \\
\hline
$b$ (arcsec)& 1.35 & 1.3446 & 0.0008 & 1.3511 & 0.0004 \\
$\epsilon$ & 0.065 & 0.0644 & 0.0011  & 0.0668 & 0.0004 \\
$\theta$ (degrees) & 0 & -0.55 & 0.32 & 0.604 & 0.14\\
$r_c$ (arcsec) & 0 & 0.0018 & 0.0080  & 0.0026 & 0.0048 \\
$x_L$ (arcsec) & 0 & 0.0402 & 0.0028 & fixed & - \\
$y_L$ (arcsec)& 0 & 0.0070 & 0.0026 & fixed & -
\end{tabular}
\caption{Mean and standard deviation of model parameters for 100 Monte-Carlo simulations based on the best-fitting PIEP model.}
\label{tab:mc_params}
\end{table}

Having shown that the $\chi^2$ surface has desirable properties, we now examine the results of the simulations.
Table \ref{tab:mc_params} shows the mean and standard deviation of best-fit parameters for the 100 simulated images.
The results with the lens centre both free and fixed are included.
The main result from the simulation is that the lens model has been recovered well but there are biases in some parameters, in the sense that the standard deviation of the fitted parameter value is less than the difference between the mean and true parameter value.
This bias is a systematic error introduced by analysing the processed radio images rather then the measured visibilities.
Fortunately, the relative bias is small: $\sim 0.5\%$ for the critical radius and $\sim 2\%$ for the ellipticity.
Compared with the error in the fitted parameters (as described in the next section), these systematic errors are very similar in magnitude, and therefore do not make the results of the analysis unusable.
It is encouraging that the lens systems can be modelled using the processed radio images with modest systematic errors. The main properties of the system: critical radius, ellipticity and position angle have been recovered. The location of the lens centre suffers from the worst systematic error but the deviation is of the order 1 pixel width (0.05 arcsec in these images).

\subsubsection{Reconstruction significance and error estimates}
\label{sec:reconerrs}

Having shown that the software can correctly reproduce the lens model parameters, this section focuses on quantifying the power of the software to distinguish between lens models, and the accuracy of the model parameter measurements.

Here we test if data with the same assumed errors and default source pixel value ($A$) can pick the correct lens model from a number of similar candidates.
The data was tested using PIEP, SPEMD and SIS+$\gamma$ models for a single noisy simulation.
The simulations were performed with an assumed pixel noise of $7.5\mu$Jy and $A$ = 0.01mJy.
For the SPEMD, two values of the mass profile where used: $\beta=0.8$ and $\beta=1.2$.

Table \ref{tab:lobe_alternate_massmodels} shows the best-fit $\chi^2$ and lens model parameters for each model.
The table shows, as expected, the incorrect mass models generate worse $\chi^2$ values. The evaluation is qualitative but supports the idea that radio images still contain much information about the lens when processed in image space.

\begin{table}
\centering
\begin{tabular}{l|c|l}
Model        & $\chi^2$ & Best parameter values \\ \hline
PIEP         & 20517 & $b=1.352$, $\epsilon=0.0669$, $\theta=0.65$ \\
SIS+$\gamma$ & 36454 & $b=1.358$, $\gamma = 0.0589$, $\theta=0.77$ \\
SPEMD        & 85155 & $b=1.446$, $q=0.867$, $\theta=4.14$, $\beta=0.8$ \\
SPEMD        & 27627 & $b=1.285$, $q=0.750$, $\theta=-2.44$, $\beta=1.2$ 
\end{tabular}
\caption{Best-fit model parameters and $\chi^2$ for a radio lobe simulation with alternate (incorrect) mass models. The SPEMD models had fixed $\beta$ and $r_c=0$.}
\label{tab:lobe_alternate_massmodels}
\end{table}

The next task in modelling the simulated lensed lobe is to quantify the uncertainty in the fitted parameters.
If the noise in the image pixels is uncorrelated, then the uncertainty in each parameter can be calculated by examining the shape of the $\chi^2$ surface around the minimum.
For radio maps, this is not immediately possible because the noise in neighbouring image pixels is correlated by the beam, hence a change in the source which affects a single pixel in the image is spread over an entire beam. In this case the total flux in a beam is 22 pixel units, so a change which generates a $\Delta \chi^2 = 1$ in a given image pixel (with no beam) makes $\Delta \chi^2 \sim 22$ over the real image (to first order).

Using Table \ref{tab:mc_params}, the standard deviation of the parameter value is interpreted as the $1\sigma$ error. With the known deviations and the well behaved properties of the $\chi^2$ surface around the minimum, one might ask whether the errors can indeed be derived directly from the $\chi^2$ surface. As in the previous section, confidence limits can be derived from the $\chi^2$ by marginalising over parameters.
The pixels are not independent in the radio map, so $\chi^2$ should be divided by the integrated beam so that it reflects contributions from individual pixels in the map.
\begin{figure}
\centering
\psfig{figure=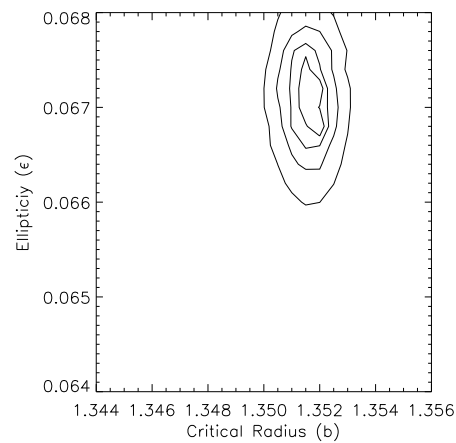,width=60mm}
\caption{Marginalised contours of $\chi^2$/beam area for the test lensed lobe for two parameters. Contours are 2.3, 4.6, 9.2 and 18.4 above the minimum.}
\label{fig:chisurf}
\end{figure}
Fig. \ref{fig:chisurf} shows the marginalised scaled ($\chi^2/22$) surface for the PIEP model with contours 2.3, 4.6, 9.2 and 18.4 above the minimum, corresponding to confidence intervals of 68, 90, 99 and 99.99 percent respectively. It is immediately apparent that the 68 percent interval corresponds very well with the $1\sigma$ errors derived from the Monte-Carlo simulations. This was true for all parameters, indicating again that the properties of the $\chi^2$ surface are suitable for error estimation.
Hence by marginalising over parameters and dividing $\chi^2$ by the total of the beam, the $\chi^2$ surface provides a good estimate of $1\sigma$ errors.
This is extremely useful and means that lengthy Monte-Carlo simulations are not necessary.

\subsubsection{Source Reconstruction}
\begin{figure*}
\mbox{
\hspace{-10mm}
    \subfigure{\psfig{figure=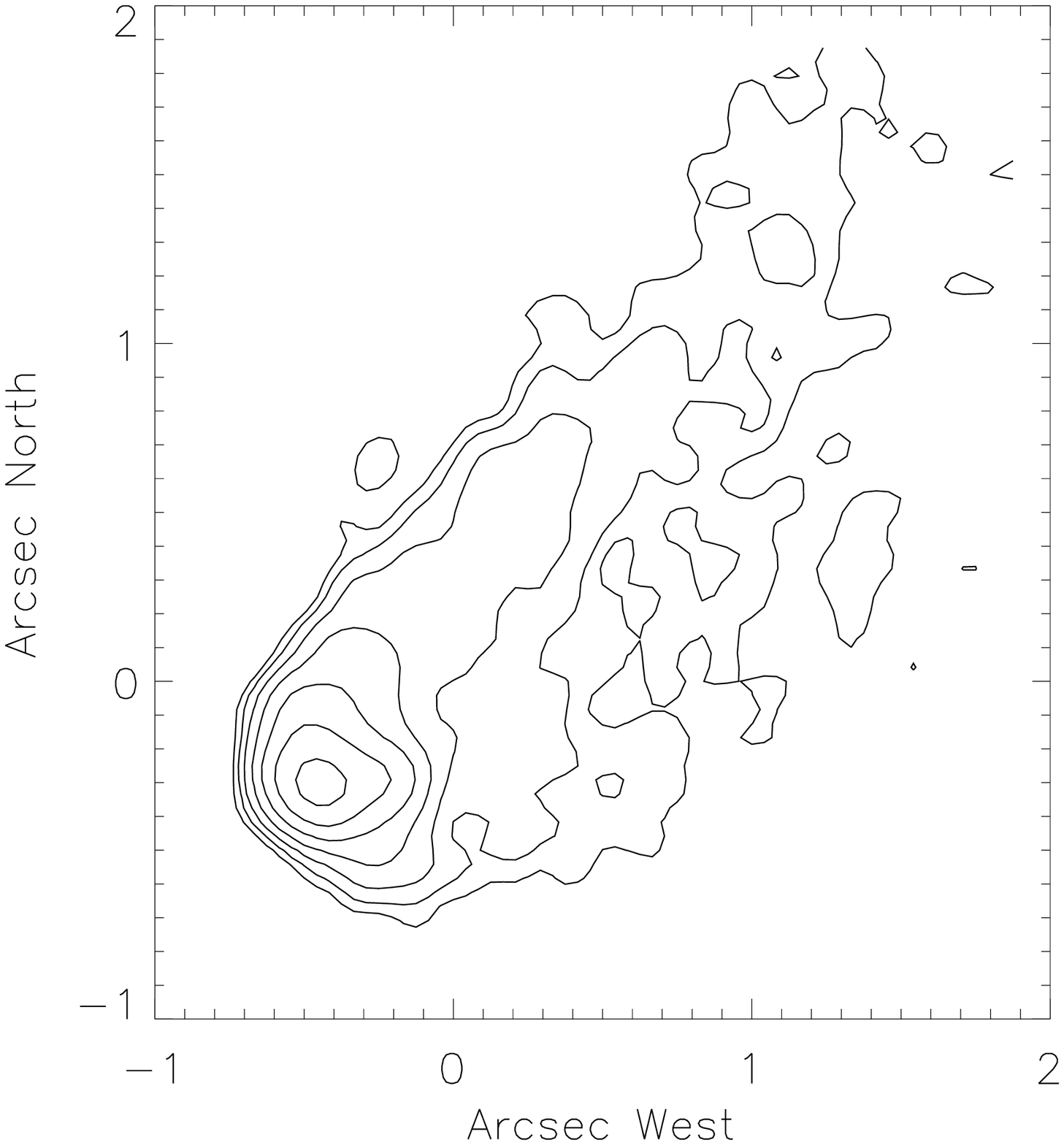,width=50mm}}
    \subfigure{\psfig{figure=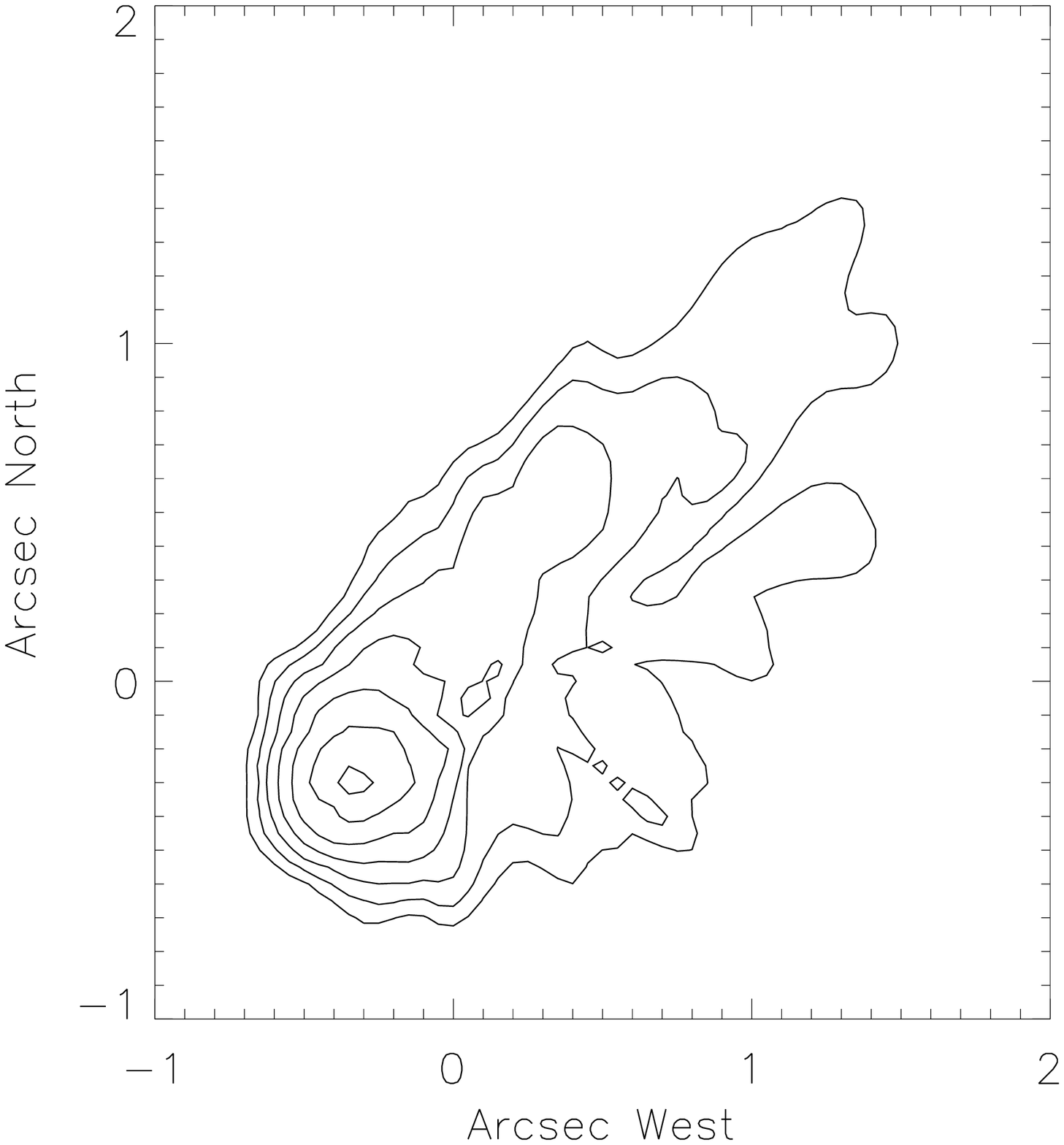,width=50mm}}
    \subfigure{\psfig{figure=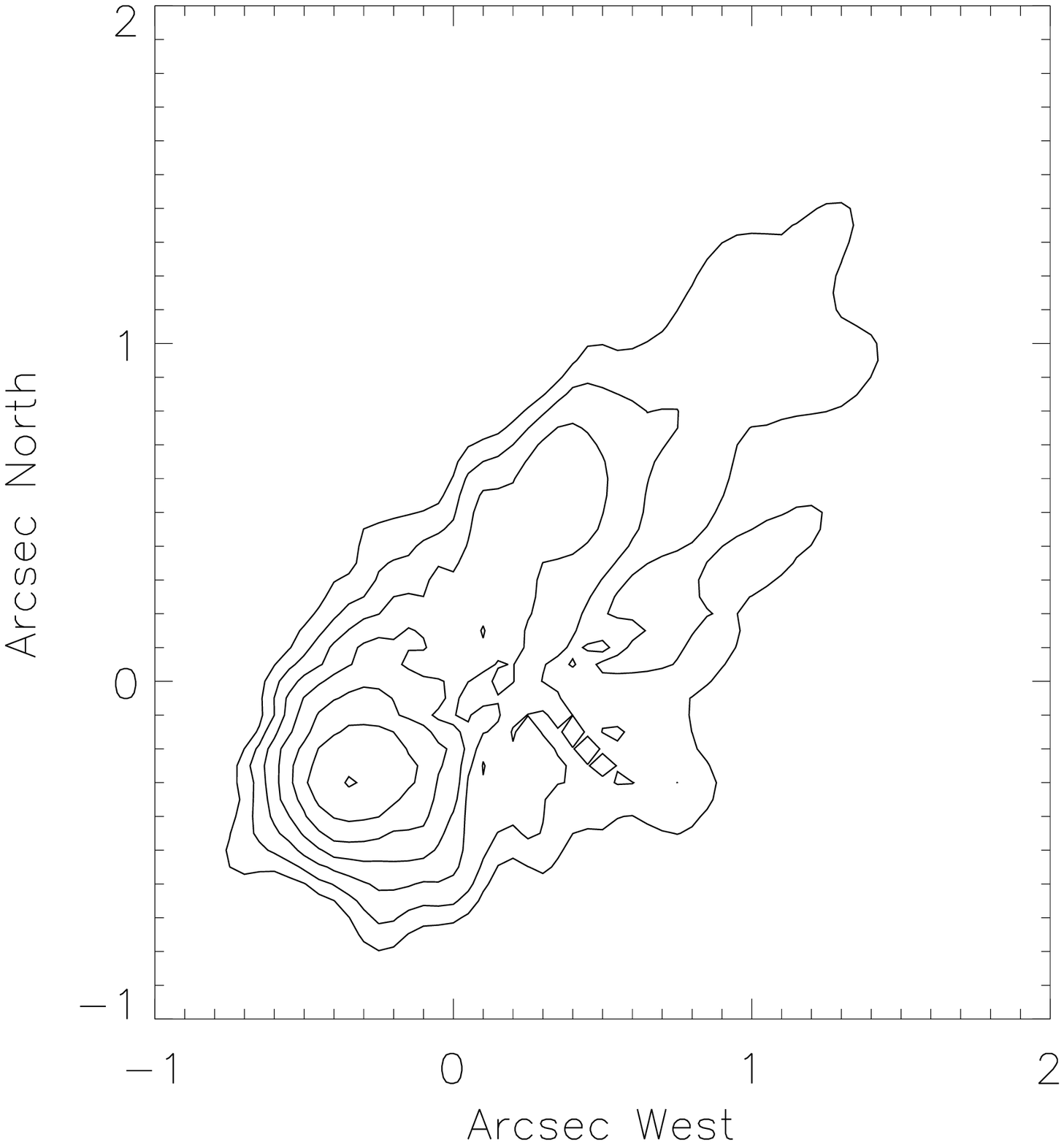,width=50mm}}
}
\caption{Reconstructions of the source for the simulated lensed lobe. Left: The original (unlensed) lobe. Centre: The idealised case where the visibilities contained no noise. The reconstruction is almost identical to the original with some low level structure around the brightest spot. Right: The reconstruction for the best-fitting PIEP model with real (noisy) data. In all images, contours double from 80 $\mu$Jy beam$^{-1}$}
\label{fig:recon_fakelobe}
\end{figure*}

In Fig. \ref{fig:recon_fakelobe} the reconstructed lobe is shown. The left image is the original unlensed lobe. The centre image shows the reconstructed lobe from a 1 hour noiseless integration.
The reconstructed lobe is very similar to the known true source even though the lensed image is a CLEANed map.
The image on the right is the reconstructed source for the best-fitting PIEP model from a 1 hour integration with noise.
Again, the lobe has been recovered quite well, with some low-level structure around the brightest spot.

\section{Examples}\label{sec:examples}

In this section we use archival HST and VLA data to model the sources HST J15433+5352 and MG1549+3047. \textsc{Lensview} has also been used in a detailed analysis \citep{2003ASPC..289..449G,2005MNRAS.360.1333W} of the optical Einstein ring ER 0047-2808 \citep{Warren1996MN,Warren1999,2003ApJ...583..606K}.

\subsection{The optical lensed arc HST J15433+5352}\label{sec:J15433}

\begin{figure}
\centering
\includegraphics[scale=0.75]{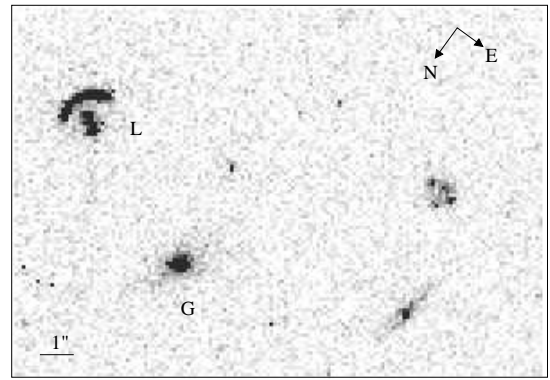}
\caption{HST J15433+5352 lens galaxy (L) and companion galaxy (G) from the HST F450W image.}
\label{fig:j15433_gals}
\end{figure}

HST J15433+5352 \citep{1999AJ....117.2010R} was one of several gravitationally lensed objects found in the HST Medium Deep Survey. The data are available (in several filters) as a WFPC2 association from the HST archive including variance files. The pixel scale of the data is 0.1 arcsec pixel$^{-1}$. In this analysis we use only the F450W image which has the best signal in the arcs and the faintest lens galaxy.
A section of the F450W image containing the lens galaxy (L) and companion galaxy (G) is shown in Fig. \ref{fig:j15433_gals}.

\begin{figure}
\centering
\includegraphics[scale=2]{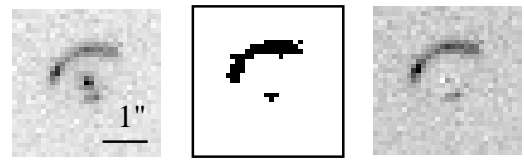}
\caption{HST J15433+5352 data. Left: The $31 \times 31$ cutout from the HST F450W data. Centre: The mask used when fitting the galaxy brightness profile. Right: the galaxy subtracted data.}
\label{fig:j15433_data}
\end{figure}

The lens galaxy was subtracted from a $31 \times 31$ section of the data using the \textsc{Galfit} code \citep{2002AJ....124..266P} v2.0.3 using a de Vaucouleurs light profile. The lensed arcs were masked to prevent them from affecting the subtraction. The data, mask and galaxy subtracted image are shown in Fig. \ref{fig:j15433_data}.

Redshifts for the lens ($z=0.497$) and source ($z=2.092$) have been measured by \citet{2004ApJ...611..739T}. The lens galaxy has a nearby (4.7 arcsec, $z=0.506$) companion and is likely part of a larger group so there will be a significant external shear in this system. The properties of the lensing galaxy and companion galaxy have recently been measured by \citet{2004ApJ...611..739T}.
In this analysis we will focus only on the properties of the overall mass distribution in the galaxy and how well it can be constrained using lensing.

We define an annulus-shaped mask of 121 pixels to encompass the lensed arcs and surrounding few pixels of the image.
Initial experiments found what seemed to be very good fits although the formal $\chi^2$ was poor. This prompted a comparison of the noise properties of the image and the variance file. Using a $50 \times 40$ square pixel section of blank sky near the lensed image, we generated a histogram of the data values. The blank sky had mean 0.06 and standard deviation 1.36 counts. The median $\sigma$ of the variance file over the same region was 0.754, indicating that the variance had been underestimated. To rectify this problem, 0.6 was added to the sigma image (i.e. the square-root of the variance) and a new variance image was generated.

To model the lens in HST J15433+5352, we again used a pseudo-isothermal elliptical potential, PIEP, with an external shear.
The centre of the lens was fixed to the centre of the light profile and we used a zero core radius.
The external shear is defined by its magnitude $\gamma$ and position angle $\theta_{\gamma}$.
There are therefore 5 parameters ($b, \epsilon, \theta_{\epsilon}, \gamma, \theta_{\gamma}$) in the lens model.
The companion galaxy, (G), and nearby group will contribute to the lensing properties of this system.
We aim to determine if the data in an extended system such as this can distinguish between lens models with and without external shear. We note also that \citet{2004ApJ...611..739T} performed an analysis of this system as part of a larger set of lenses using a completely different method. We compare results below.

Notwithstanding the discussion about degrees of freedom in the source model in \S\ref{sec:targ_chisqu}, the image plane mask of 121 pixels does not leave many degrees of freedom in the source before the target  $\chi^2$ becomes zero or negative.
Initial experiments showed that source plane pixels had to be at most $1/2$ the size of image pixels and that the source is elongated.
We chose the smallest possible source plane that could accommodate the reconstructed source with a source-to-image plane pixel scale ratio of 1/2.
The final source plane is $10 \times 10$ pixels and offset from the centre of the lens plane.
Using equation \ref{eq:my_dof}, the target $\chi^2$ is $18 \pm 6$.
\begin{table}
\centering
\begin{tabular}{l|ll}
Model           & $\chi^2$ & params  \\ \hline
PIEP            & 75.9 & $b = 0.525 \pm 0.015$ \\
                &      & $\epsilon = 0.16 \pm 0.01$ \\
                &      & $\theta_{\epsilon} = -120 \pm 3$ \\
PIEP + $\gamma$ & 73.1 & $b = 0.525 $ \\
                &      & $\epsilon = 0.11$ \\
                &      & $\theta_{\epsilon} = -134$ \\
                &      & $\gamma = 0.06$ \\
                &      & $\theta_{\gamma} = -104 $ \\
\end{tabular}
\caption{Results for simple mass models of HST J15433+5352. For the PIEP + $\gamma$ model, there were many combinations of $\epsilon,\theta_{\epsilon},\gamma,\theta_{\gamma}$ that yield very similar $\chi^2$, but none that were a substantial improvement on the PIEP model. For this reason, no errors are given.}
\label{tab:j15433_res}
\end{table}

\subsubsection{Results}
\begin{figure}
\centering
\includegraphics[scale=0.4]{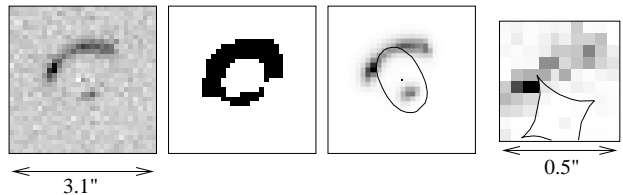}
\caption{HST J15433+5352 models. Left: the galaxy subtracted data. Centre left: The mask of 121 pixels used in the modelling. Centre right: The best-fit PIEP model image and critical line. Right: The model source and caustic.}
\label{fig:j15433_models}
\end{figure}

Best fitting parameters of the modelling are shown in Table \ref{tab:j15433_res}.
The significance of the $\chi^2$ values is discussed in the next section.
The surprising result from the modelling is that a purely elliptical model is equally able to reproduce the data as a model which includes shear. Put another way, the combined effect of the external perturbations plus intrinsic ellipticity in the lensing galaxy is indistinguishable from a different, purely intrinsic, ellipticity (and orientation) in the galaxy
itself (at least in this lens). This result is in contrast to other results \citep[e.g.][]{1997ApJ...482..604K}, which find that many QSO lenses with only positional information can only be explained with the presence of an external shear.
A possible reason why an external shear is not required, at least in this lens, is that the lens is essentially a cusp image. The reconstructed source, shown in Fig. \ref{fig:j15433_models}, lies across a cusp. The lensing properties of cusps have been shown to be quite generic \citep{1992grle.book.....S,1992A&A...260....1S} so the section of the arc that lies very close to the critical line will be locally independent of the particular details of the lens model. The section of
arc that is not close to the critical line has its shape determined both by the lens model and the shape of the source, so it cannot reveal anything conclusive about the lens without prior knowledge of the source properties. It is reassuring, however, that a simple lens model and morphologically uncomplicated source can explain the image.

Errors on the fitted parameters are also shown in Table \ref{tab:j15433_res} for the PIEP model. The errors were generated using the properties of the $\chi^2$ surface around the minimum.
For the PIEP + $\gamma$ model, the ellipticity and orientation angle parameters $\epsilon,\theta_{\epsilon},\gamma,\theta_{\gamma}$ are virtually degenerate hence no errors are given. Although our modelling found a `best-fit' for this model,
many combinations of parameters produced $\chi^2$ values which were within $\Delta \chi^2 \sim3$ of the best fit, therefore equally as good. For any (reasonable) set of values ($\epsilon$, $\theta_{\epsilon}$) there exists ($\gamma,\theta_{\gamma}$) such that the $\chi^2$ is as good, but not significantly better, than the PIEP.

The critical radius, $b$, differs from \citet{2004ApJ...611..739T} (0.36 arcsec), however they modelled the companion galaxy, G, as an isothermal sphere. Given its close proximity to the lens galaxy, L, the halo of G will contribute substantial convergence to their lens model, so their smaller Einstein radius is merely making up the difference. The critical radius is also substantially different from the value found by \citet{2001AJ....122..103K} ($b=0.58$ arcsec) who used simple elliptical isothermal models like those used here.
This systematic difference is huge (10\%!) compared to the random error from the fitted models.

The model image and reconstructed source for the PIEP model is shown in Fig. \ref{fig:j15433_models}. The PIEP + $\gamma$ model looks identical, so it is not shown. Our model shows a source, approximately 0.5 arcsec in length, which straddles a cusp. The brightest region in the source is \emph{outside} the caustic. The source reconstruction is qualitatively similar to \citet{2004ApJ...611..739T} and \citet{2001AJ....122..103K} but differs in detail. Comparing to \citet{2004ApJ...611..739T}, their source has the brightest region outside the caustic but the model is curved around the cusp. Compared to \citet{2001AJ....122..103K}, their source is placed at a different location, has a different position angle relative to the caustic and is more round.
The model image generated by their source has significant residuals as seen in fig. 4 of \citet{2001AJ....122..103K}. In
particular, the fainter part of the arc is too bright and the brightest region of the arc (at top and left of Fig. \ref{fig:j15433_models}, LHS) are too faint. Given a parametric source model, however, it is the best that can be done.
Hence, the use of a parametric source by \citet{2001AJ....122..103K} has resulted in a significant systematic error in the critical radius of their best lens model.

\subsubsection{Source plane pixel thresholding}
At face value, neither of the models produce a statistically acceptable result. The reason for this is obviously the large source plane and the number of degrees of freedom used in the model. From Fig. \ref{fig:j15433_models} it is clear that the source pixel threshold method is appropriate for this model.

\begin{table}
\centering
\begin{tabular}{cccc}
Threshold & Number of & Adjusted & \\
 T &  zeroed pixels &  target $\chi^2$ & New $\chi^2$ \\ \hline
1.0 & 30 & $48\pm 10$ & 74.8 \\
1.5 & 40 & $58\pm 11$ & 75.5 \\
2.0 & 46 & $64\pm 11$ & 76.9 \\
2.5 & 50 & $68\pm 12$ & 78.9 \\
3.0 & 55 & $74\pm 12$ & 82.8 
\end{tabular}
\caption{Results of adjusting the target $\chi^2$ and source for the PIEP model based on a threshold, T, below which source pixels are set to zero. In this example, the noise in each pixel is approximately 1.4 counts and the default pixel value parameter, $A$, is 2.0. For each threshold level, the new source was re-projected through the best-fitting model and a new $\chi^2$ calculated. This table shows setting pixels with value $< 2.5$ to zero makes no difference to the model.}
\label{tab:j15433_adjustchi}
\end{table}

\begin{figure}
\centering
\includegraphics[scale=0.25]{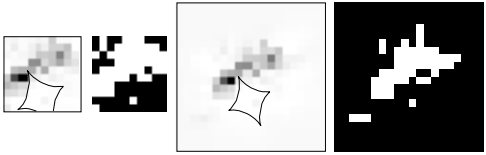}
\caption{HST J15433+5352 source models and source threshold masks.}
\label{fig:j15433_srcs}
\end{figure}

Using this scheme, we must find T for our source. This is straightforward and we find setting T=2.5 makes 50 pixels go to zero and produces a $\Delta \chi^2$ of 3.0 as shown in Table \ref{tab:j15433_adjustchi}. The pixels set to zero are shown in Fig. \ref{fig:j15433_srcs}. Thus, our revised source plane has 50 degrees of freedom so the target $\chi^2$ is $68 \pm 12$. Using this revised definition, the original models are acceptable. For comparison, the typical noise level in the image is 1.4 counts, the default source pixel value parameter, $A$, is 2.0 and the peak count in the reconstructed source is 77.0. Hence we really are zeroing only those pixels which have not contributed anything to the model.

Our method shows that the combination of using a mask on the image plus a source plane zeroing threshold leads to a sensible and achievable definition for the target $\chi^2$. Formally, we are simply replacing $N_{\rm{source}}$ in (\ref{eq:my_dof}) with $N_{\rm{source}} > T$.

\subsection{MG1549+3047}
MG1549+3047 (also known as 4C+30.29) is a $z=1.17$ radio lobe being lensed by a $z=0.11$ SB0 galaxy \citep{1993AJ....105..847L,1996AJ....111.1812L,2003MNRAS.343L..29T}.
\label{sec:data_reduction}
We use previously unpublished, archival VLA
data of MG1549+3047 in this analysis.
\citet{1993AJ....105..847L} used VLA data taken on 1991 August 2 in A configuration with a 20 min integration and achieved an RMS noise of 0.01 mJy beam$^{-1}$. MG1549+3047 was re-observed on 1992 November 18 at 1.46 GHz (5 min), 4.89 GHz (30 min), 8.41 GHz (50 min) and 14.96 GHz (190 min) again with the VLA A configuration. We use this new 8.41 GHz data in this work because the 14.96 GHz data has very poor signal in the ring. Three antennas (10,27,29) were off-line during the observations.

We followed a standard VLA calibration procedure using AIPS and calculated 3C286 fluxes using the AIPS `setjy' task.
The data were quite clean and we used one round of self-calibration to make an image. The beam size was $0.21 \times 0.24$ arcsec. 
The off-source RMS noise in the map is 0.041 mJy beam$^{-1}$ which is almost twice the theoretical noise of 0.025 mJy beam$^{-1}$and four times the RMS noise stated in \citet{1993AJ....105..847L} table 2.
Since fig. 2 in \citet{1993AJ....105..847L} shows contours doubling from 0.2 mJy  beam$^{-1}$,
there may have been a typographical error and the actual noise in their map was 0.13 mJy beam$^{-1}$ rather than the stated value of 0.013 mJy beam$^{-1}$.
The larger value would also be much more consistent with the theoretical value of 0.05 mJy beam$^{-1}$ for a 20 min integration.

\begin{figure*}
\hspace{-5mm}
\subfigure{\psfig{figure=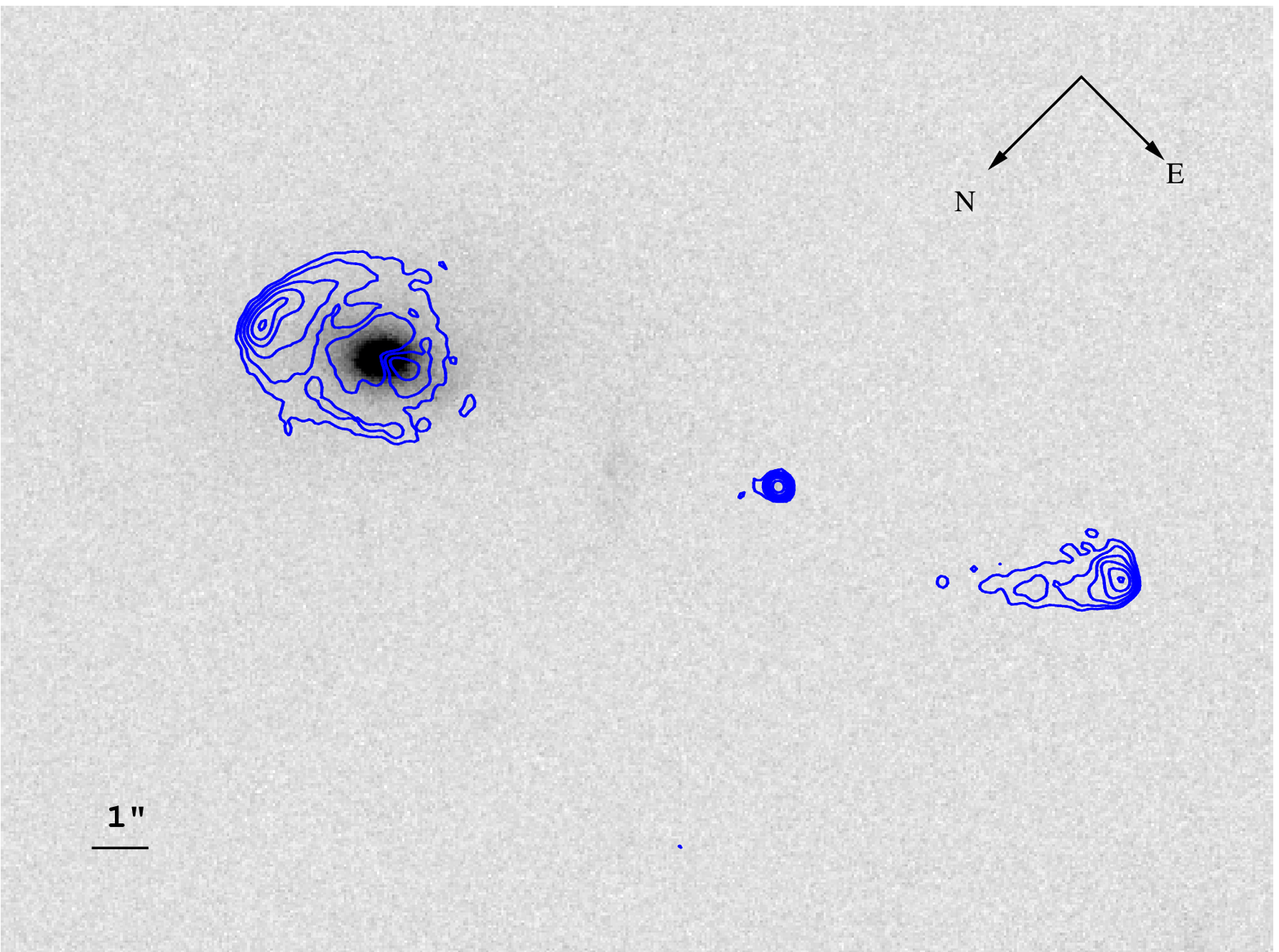,width=80mm}} \quad
\subfigure{\psfig{figure=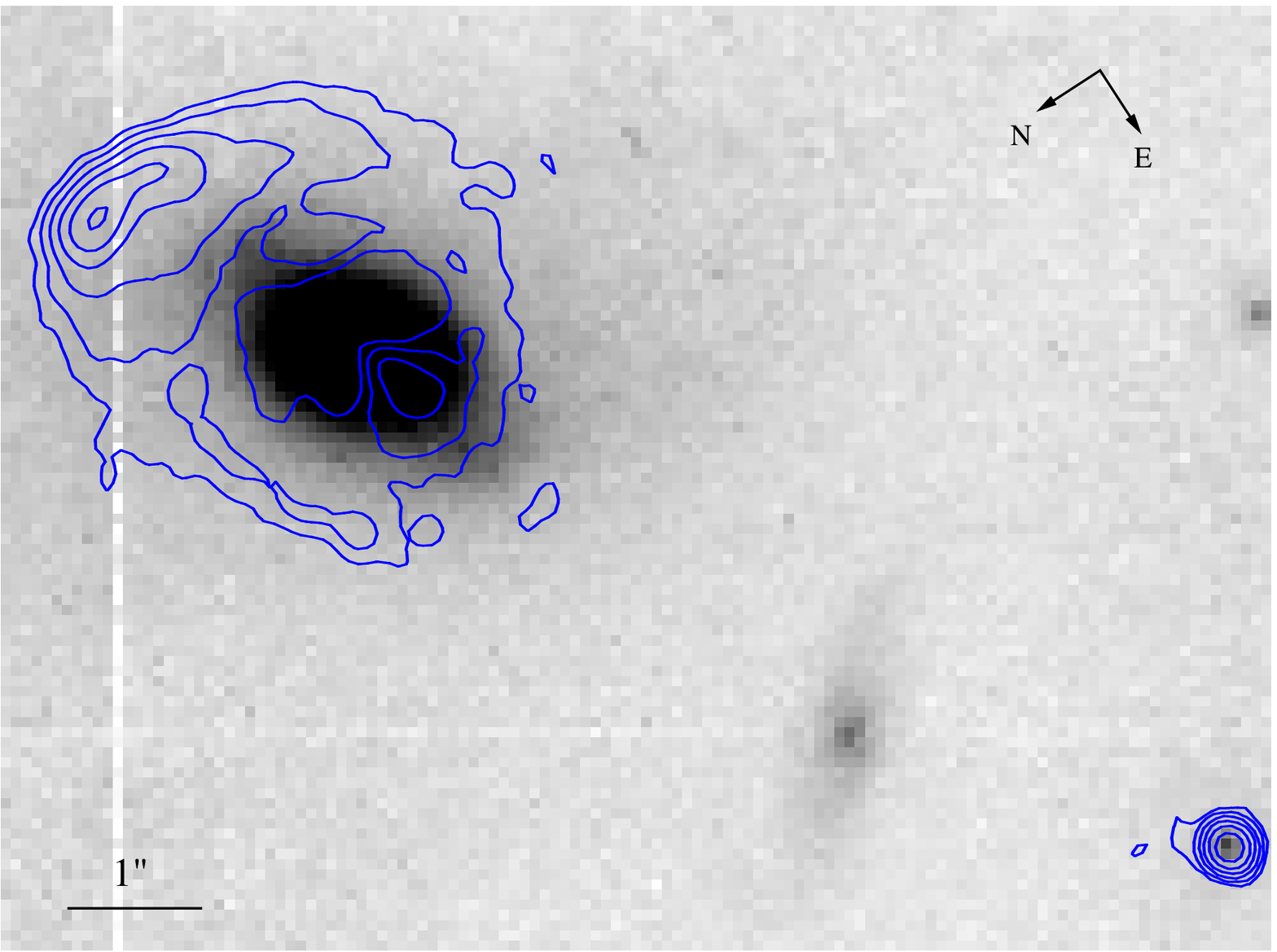,width=80mm}}
\caption{MG1549+3047 radio galaxy and lobes. Left: 8.4 GHz radio contours are overlaid on the HST WFPC2 F814W image. Right: 8.4 GHz radio contours are overlaid on the HST NICMOS F160W image. Radio contours double from 0.2mJy. Between the radio core and lens is an intermediate redshift (z=0.604) galaxy.}
\label{fig:contours_on_hst}
\end{figure*}

HST NICMOS F160W and WFPC2 image associations (proposal id: 7495, PI: Falco) were obtained from the HST archive which show the faint $z=1.17$ radio galaxy, which is the source of the lobes plus a nearby $z=0.604$ galaxy \citep{2003MNRAS.343L..29T}.
We used the HST images purely for astrometric purposes and performed no processing other than
that which is automated by the archive pipeline (flat fields, CR rejection, image combining).
We aligned the radio and optical images using the radio galaxy with a combined positional uncertainty of 0.04 arcsec.
The combined radio and optical images are shown in Fig. \ref{fig:contours_on_hst}.
We calculate the centre of the lens galaxy to lie at (J2000) RA: $15^{h} 49^{m} 12\fs330$, DEC: $30\degr 47\arcmin 16\farcs51$ with an uncertainty of 0.04 arcsec.
The ring has lensed flux almost all the way to the centre of the galaxy, so it is extremely important that the location of the lens centre is known accurately. If the location of the lens centre is incorrect, any measurement of the properties of the mass in centre of the galaxy will be affected.

We model MG1549+3047 using a simple PIEP  lens model to compare to previous work and derive some basic properties of the lens.
\citet{1993AJ....105..847L} presented a lens model based on matching various features of the lensed image, then performing a QSO-like analysis using the locations of the features. Their lensing potential,\
$\psi = \beta \left( \left[ 1 + ( 1- \epsilon) (x/s)^2 + (1 + \epsilon)(y/s)^2 \right]^{1/2} - 1 \right) $
has the same deflection angle properties as the PIEP with $b=\beta/s$ and $r_c = s$.

\begin{figure*}
\hspace{-5mm}
\subfigure{\psfig{figure=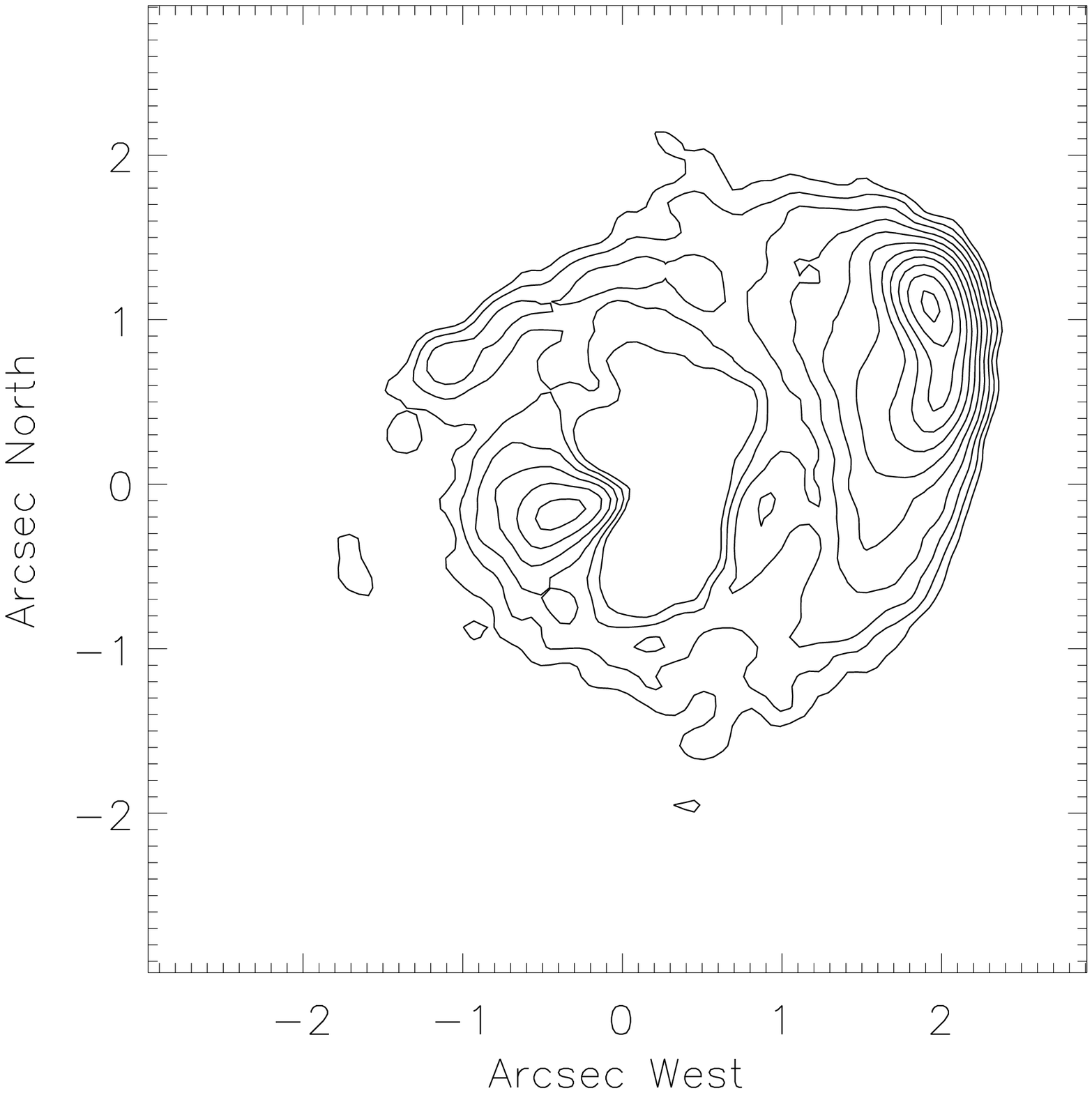,width=58mm}}
\subfigure{\psfig{figure=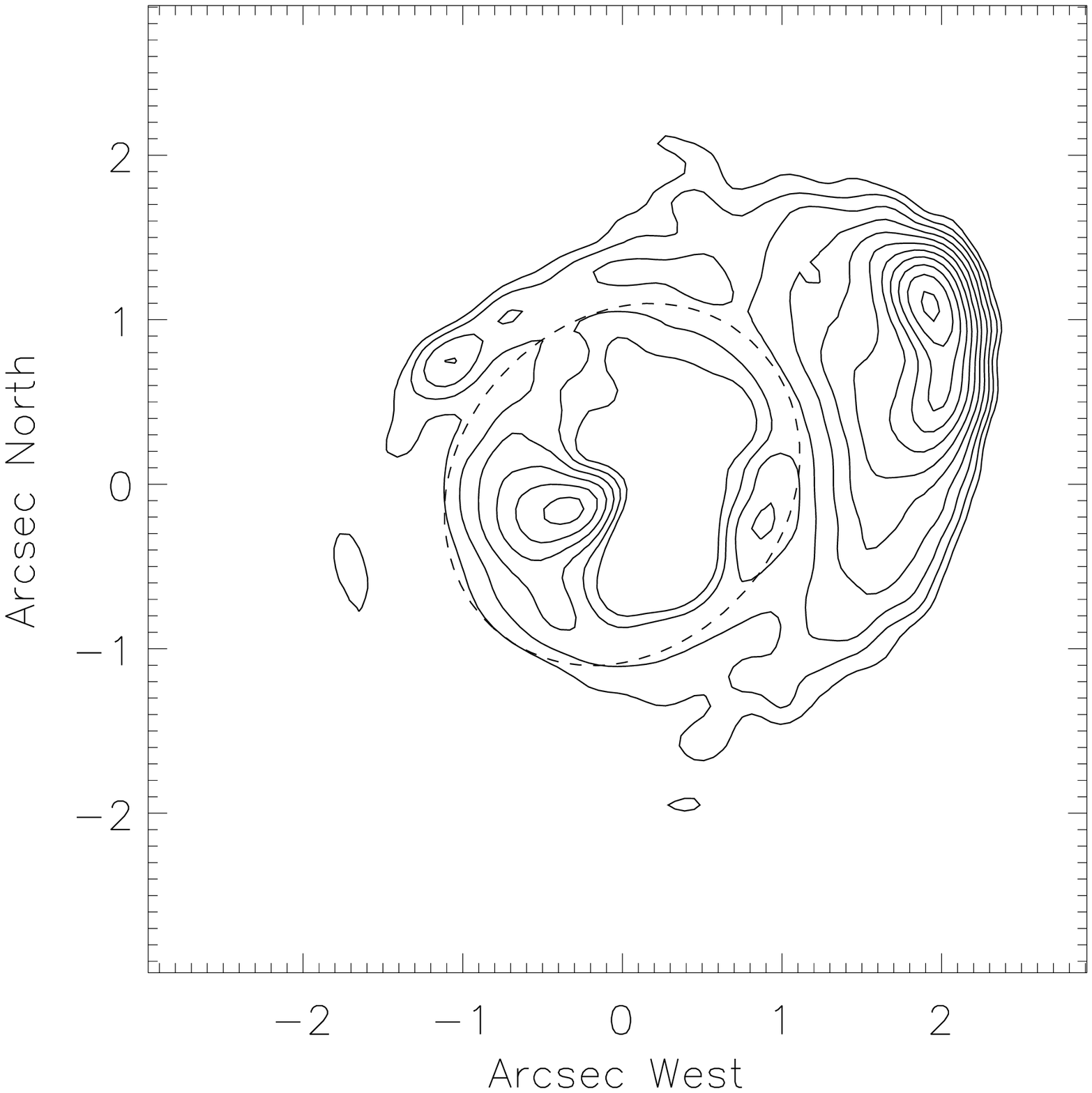,width=58mm}}
\subfigure{\psfig{figure=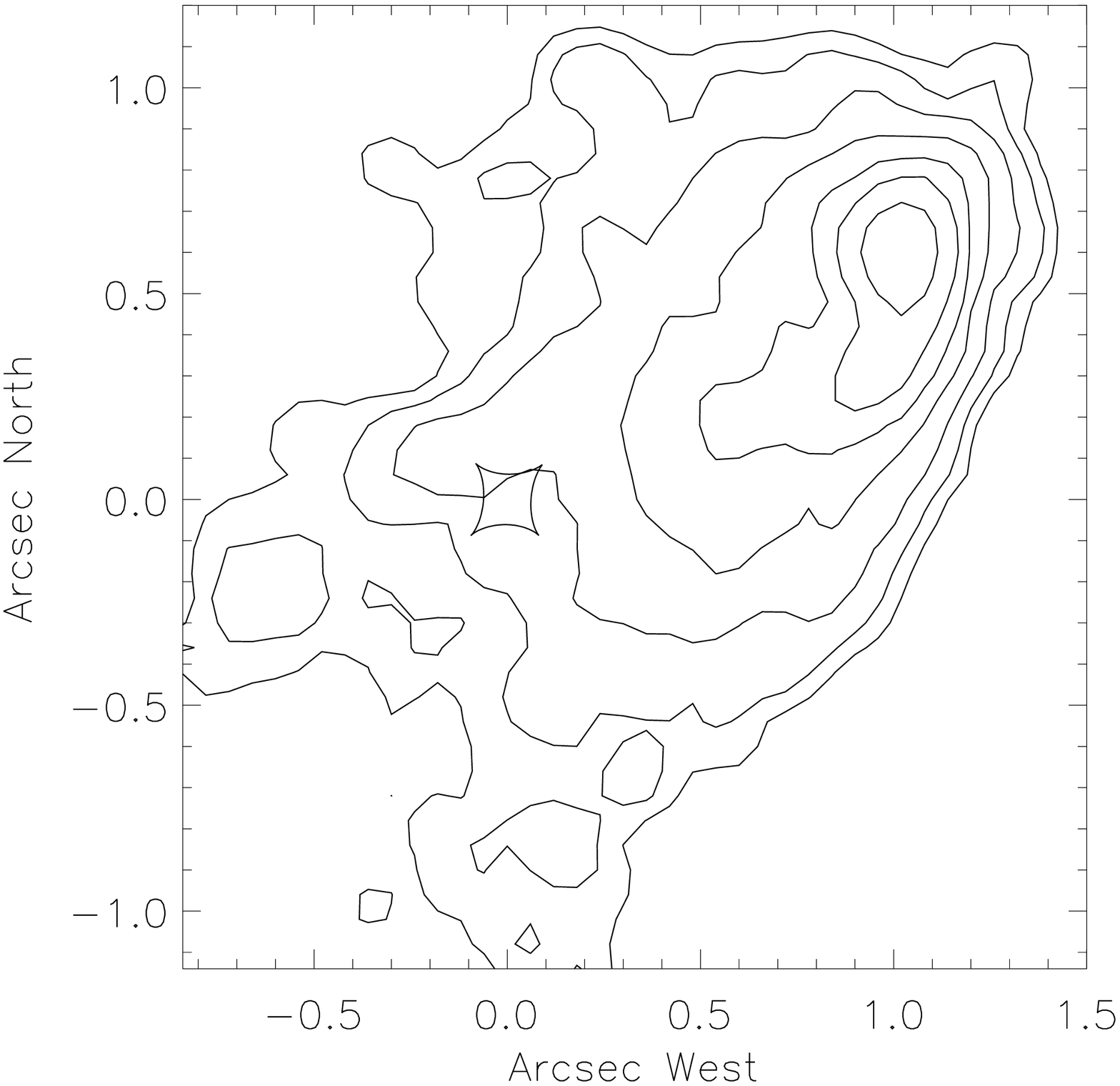,width=58mm}}
\caption{MG1549+3047 data and models. North is up, East is left and position (0,0) represents the centre of the lens. Left: The lensed lobe. Centre: Model PIEP image with tangential critical line (dashed).
Contours increase by $\sqrt{2}$ from 0.2mJy beam$^{-1}$ for the data and model.
Right: The reconstructed source after smoothing with a boxcar filter of width 3. The caustic is also shown.
Contours for the source increase by 2 from 0.08 mJy pixel$^{-1}$}
\label{fig:mg1549_data}
\end{figure*}

We searched parameter space using the results from \citet{1993AJ....105..847L} as a starting point.
The parameters were unrestricted except for the lens centre, which we allowed to vary by up to 0.1 arcsec from our measured position.
A $99 \times 99$ grid of pixels from the radio data was used for the modelling, shown in Fig. \ref{fig:mg1549_data} (LHS). The variance in each pixel was defined to be $(0.04)^2$, i.e. the square of the off-source RMS noise in the map.

\begin{table}
\begin{tabular}{l|r|l|r}
 & This & & L93 \\
Parameter & work &  & result \\ \hline
b (arcsec) & 1.105 &$\pm 0.005$ & 1.15 \\
$\epsilon$ & 0.056 &$\pm 0.006$ & 0.073\\
$\theta_{\epsilon}$ (deg E of N) & -47 &$\pm$ 1 & -48 \\
$s$,$r_c$ (arcsec) & 0.132 & $\pm 0.03$ & 0.22 \\
$x_L$ (arcsec E of radio core) & -6.59 & $\pm 0.02$ & -6.64 \\
$y_L$ (arcsec N of radio core) & 3.33 & $\pm 0.02$ & 3.38 \\
Mass enclosed ($\times 10^{10} M_{\odot}$): & 7.06 &$\pm 0.06$ & 7.72 \\
$\sigma_v$ (for best fit b) (km s$^{-1}$) & 209 & &213
\end{tabular}
\caption{Best fit PIEP model parameters for MG1549+3047 and corresponding values from \citet{1993AJ....105..847L}, converted to our cosmology.}
\label{tab:mg1549_params}
\end{table}

The result of our modelling, along with some results from table 4 in \citet{1993AJ....105..847L} are shown in Table \ref{tab:mg1549_params}. The errors were calculated as discussed in \S \ref{sec:reconerrs}.
The most significant difference is that of the Einstein radius. Since the mass enclosed within the ring is $\propto b^2$, the difference in Einstein radius changes the mass by 10 percent, which is quite a substantial amount given the accuracy to which Einstein radii are usually measured. Our predicted velocity dispersion, 209 km s$^{-1}$, is slightly lower, but consistent with, the measured value of 227 $\pm 18$ km s$^{-1}$ \citep{1996AJ....111.1812L}.

The lens centre was found to be very close (0.024 arcsec) to the position predicted by the combined optical/radio astrometry. The best fitting lens centre is at RA: $15^{h} 49^{m} 12\fs330$, DEC: $30\degr 47\arcmin 16\farcs42$. The uncertainty on the fitted lens centre (0.02 arcsec in each direction) is smaller than the uncertainty from the combined astrometry. This result shows that the software can be used to find the centre of the lens in cases where there is some uncertainty. We found that the best-fitting  model core radius, $s$, was affected by the position of the lens centre. This can be understood by noting that as the lens centre moves East of its known position, a radial critical line is no longer required to truncate the image. Rather, the image is simply disappearing at the centre of the lens. Despite the correlation between parameters, the lens centre was unambiguously identified by the modelling.

\begin{figure}
\centering
\psfig{figure=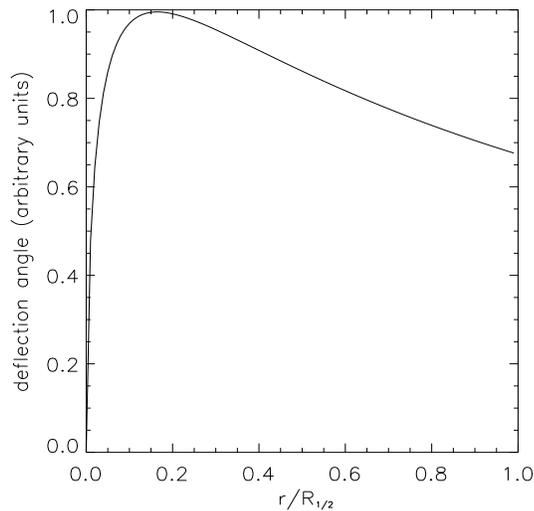,width=75mm,height=70mm}
\caption{Relative magnitude of the deflection angle for a de Vaucouleurs model with the radius in units of the galaxy 1/2 light radius, $R_{1/2}$.}
\label{fig:devauc_circ_defl}
\end{figure}

Another noteworthy property of the model is the requirement of a small constant surface-density core.
The lensing galaxy has a prominent bulge which should follow an $r^{1/4}$-type light profile. (We leave a detailed analysis of the properties of the lens galaxy to a future work.) Such a profile has well defined lensing deflection properties. The magnitude of the deflection rises from zero in the centre of the galaxy, reaches a peak around 0.2 scale radii then falls away roughly as $1/r$, as shown in Fig. \ref{fig:devauc_circ_defl}. Any bulge-like galaxy's gravitational and lensing potential should be dominated by stars in the centre of the galaxy, so one would expect the lensed image to reflect the lensing properties of the bulge in the central regions of the galaxy. Further out, the galaxy's dark matter becomes more and more important so the lensing properties follow the familiar isothermal model.
Our model confirms this expectation with the requirement of a core (non-isothermal) region with zero/small deflection angles.

The reconstructed source, shown in Fig. \ref{fig:mg1549_data}, looks just like a regular FR-I type radio lobe. It has many similarities with its unlensed counterpart, including a hot-spot, diffuse tail and some slightly brighter structures in the tail. The total flux in the reconstructed lobe is 35mJy compared to 44mJy in the unlensed lobe. The overall magnification for the lobe is modest: about 3.4, but this is because the bright spot in the lobe is outside the multiple imaging region of the lens.
It is the much fainter tail which is the most highly magnified.

The significant difference in Einstein radius between our model and L93 highlights the importance of using detailed models for resolved lenses. The analysis of L93, based on identifying features in the lens, is dangerous- especially when identifying features close to the centre of the lens in low magnification regions. L93 identified three sets of features (A1,A2,B1,B2,C1,C2) in the lens. Our model shows that A2 and B2 are features which have several beam areas worth of source brightness information packed into one or two beam areas in the image. Since the source brightness varies rapidly with position in the lobe, the resulting brightness at position A2 and B2 cannot be directly mapped to another region of equal brightness in the image. A1 and B1 are a good approximation of the true surface brightness in a particular part of the source, whereas A2 and B2 are the average brightnesses of a large portion of the source spanning several beam areas.

\section{Discussion and conclusions}
\label{sec:conclusions}
Resolved gravitationally lensed images provide the power to discriminate between mass models for the lensing galaxy. However, until recently this potential has been largely untapped.
There is now a growing set of algorithms and tools for modelling resolved lensed images.
In this section we discuss issues relating to our software, \textsc{Lensview} and how it compares to other algorithms.

\emph{Mapping matrix vs non mapping matrix:} The mapping matrix has many convenient properties, not the least of which is that it correctly preserves surface brightness when projecting between the source and image planes. \citet{2003ApJ...590..673W} use ray-shooting to map source pixels to image pixels. This leads to the question: if a projected image pixel overlaps with several source pixels, which one do you choose?
This is most important for image pixels which are in a medium to low magnification region. Such pixels will overlap with several source pixels and their true value has a contribution from all. By ray shooting, only a single pixel is sampled. The effects of this sampling are unknown at present as are the effects of the simple interpolation scheme proposed by \citet{2004ApJ...611..739T}. In principle, there is no reason why a mapping matrix cannot be incorporated into the semilinear method, but the increased source-to-image pixel relationships of the mapping matrix means the matrix be to inverted by the semilinear method rapidly loses its sparseness. Obviously a mapping matrix is not useful for \textsc{LensCLEAN} which uses point sources, although the \textsc{LenTil} algorithm \citep{2004MNRAS.349....1W} uses something similar to find image positions.

\emph{Iterative vs non iterative:} The source model for our method builds up through progressive iterations of forward and reverse projection. The semilinear method calculates the source model in a single matrix inversion where the matrix has size $N \times N$ (with $N$ pixels in the image) and is presumably fairly sparse. The `sparseness' of the matrix must be reduced with an increasingly broad PSF. This will affect running times and possibly be susceptible to accumulated numerical errors. The iterative approach is not significantly affected by the size of the PSF and numerical errors do not accumulate between iterations. The time to project between source and image planes, however, is affected by the fraction of the image plane which covered by the source plane. Nevertheless, the majority of the computational time in our method is taken convolving with the PSF. Typically 100-200 iterations are required.

\emph{Running times:} Code execution times have not been given for other methods. As mentioned above, \textsc{Lensview} spends most of its time performing convolutions via the FFT. Hence running times are roughly $O(N \log N)$ where $N$ is the number of pixels in the image. For the examples in this paper, running on a 2.0GHz Pentium computer, the inner loop is completed in less than one second for J15433 and in approximately seven seconds for MG1549. Hence, large sections of parameter space can be searched in reasonable times with \textsc{Lensview}.

\emph{Entropy/regularisation:} The entropy constraint is analogous to regularisation of semilinear inversion although it should be emphasised that the way in which the source is reconstructed in each method is quite different. Any form of regularisation introduces bias into the source, however that is the price one must pay for using real (noisy) images since all image deconvolution methods will require regularisation in some form. The entropy constraint is an excellent form of regularisation because it smooths source pixels which have little or no data constraining them and enforces positivity in the source. Other choices for a source constraint are possible, but we have not explored them here.

\emph{Images vs visibilities:} Although \textsc{Lensview} currently works only with images, the extension to use radio visibilities is conceptually straightforward. It is difficult to see how the semilinear method could be extended to use radio visibilities because the matrix inversion would become impractically large- this is exactly why CLEAN and similar algorithms are used to process radio images.

A simple estimate shows that processing the data in image space provides approximately two orders of magnitude increase in the speed of analysis.
Fourier transforming between the visibilities and dirty images used in the simulations takes $\sim 1$ second. Iterative methods must be used with this volume of data, and our models (and those of SB84) usually converge in 100--200 iterations. Assuming no other computational burden, the forward and backward Fourier transforms alone will take several minutes for a single lens model for the inner loop.
Conversely, by using the CLEANed image, just the region of interest can be extracted from the larger image for use with \textsc{Lensview}.

The tests presented here show that the accuracy of the lens model parameters is biased by working in image space, however the systematic errors are small enough that the results are still useful for some applications. \citet{1996ApJ...464..556E} found similar biases after comparing results of image-space models with visibility-space models. They therefore concluded that using visibilities is preferable to images. \citet{2004MNRAS.349....1W} subsequently showed that further improvements to LensCLEAN were possible. Since \citet{1996ApJ...464..556E} did not perform any simulations, it is not known whether their visibility-based results were actually affected by the problems identified by \citet{2004MNRAS.349....1W}. Hence it is possible that systematic effects were still present.

A major advantage of using visibilities is that a well defined target $\chi^2$ exists.
In some rare cases, e.g. MG1131+0456 \citep{1988Natur.333..537H}, the entire image generated by the interferometer is contained in a very small area --- just a few square arcseconds. In that case it would be possible to process the \emph{dirty} image directly using the dirty beam as a PSF. Since there would be no artifacts due to image processing in that case,
the noise properties and target $\chi^2$ of the image would be well defined.

\emph{Mass distributions:} \textsc{Lensview} can work with an arbitrary lens mass model with negligible extra computational cost. This appears to be true also for the semilinear method, but is not so for \textsc{LensCLEAN} which requires solutions to the lens equation for each point source component. Hence \textsc{LensCLEAN} is substantially slower when using a non-analytic lens model.

The potential applications of \textsc{Lensview} extend well beyond the simple examples shown in this paper. Some of the most exciting applications are: 1) determining the slope of the mass profile in lens galaxies. Interpretation of $H_0$ measurements rely on this knowledge. \citet{2005MNRAS.360.1333W} have shown that the optical Einstein ring ER 0047-2808 has a well determined mass profile in the region of the images. 2) properties of the dark haloes of lens galaxies. It is straightforward to model a lens as a stellar (baryonic) component with a dark matter halo. Extended images will allow measurements of the properties of the halo \citep[e.g.][]{2003ASPC..289..449G}. 3) determination of the lens centre. Again, required for $H_0$ measurements.

We have presented simple elliptical isothermal lens models for the optical lensed arc HST J15433+5352 and the radio Einstein ring MG1549+3047.
Our models show that for HST J15433+5352, the data are unable to distinguish between an elliptical model and an elliptical model with external shear.
We speculate that this because the source is lying over a cusp, therefore has a very generic image configuration.
For MG1549+3047, we show that the previous lens model had an incorrect Einstein radius and therefore the mass enclosed in the ring was overestimated.
In addition, we showed that the model requires a small non-singular core, consistent with expectations for a star-dominated mass distribution in the centre of the lens galaxy.

More detailed applications of the software are possible. We look forward to a new era of precision modelling of resolved images.

\section*{Acknowledgements}

Some of the data presented in this paper were obtained from the Multimission Archive at the Space Telescope Science Institute (MAST).
STScI is operated by the Association of Universities for Research in Astronomy, Inc., under NASA contract NAS5-26555.
Support for MAST for non-HST data is provided by the NASA Office of Space Science via grant NAG5-7584 and by other grants and contracts.

The National Radio Astronomy Observatory is a facility of the National Science Foundation operated under cooperative agreement by Associated Universities, Inc.

We thank Geraint Lewis, Steve Warren, Paul Hewett and Cathryn Trott for extensive discussions and to the referee for valuable comments.
\newcommand{\apj}{ApJ}
\newcommand{\nat}{Nat}
\newcommand{\mnras}{MNRAS}
\newcommand{\aj}{AJ}
\newcommand{\pasp}{PASP}
\newcommand{\aap}{A\&A}
\newcommand{\apjl}{ApJ}
\newcommand{\apjs}{ApJS}
\newcommand{\aaps}{A\&AS}
\bibliographystyle{mn2e}
\bibliography{references}
\label{lastpage}
\end{document}